\documentclass[useAMS,usenatbib,babel]{mnras}
\usepackage{times}

\usepackage[english,english]{babel}
\usepackage{amsmath}
\usepackage{amssymb,amsfonts,textcomp}
\usepackage{array}
\usepackage{supertabular}
\usepackage{placeins}
\usepackage{hhline}
\usepackage[usenames]{color}
\usepackage{soul}
\usepackage{multirow}
\usepackage{textcomp}
\usepackage{graphicx}
\usepackage{float}
\usepackage{dcolumn}
\usepackage{bm}
\usepackage{comment}
\usepackage{color}

\usepackage{hyperref}
 \usepackage{float}
 \usepackage{balance}
\usepackage[caption=false]{subfig}

\newcommand{\mP}{{\cal P}}

\newcommand{\vr}{\textbf{r}}
\newcommand{\vx}{\textbf{x}}

\newcommand{\vk}{\textbf{k}}

\newcommand{\dd}{\hbox{d}}

\newcommand{\lin}{{\rm lin}}

\newcommand\T{\rule{0pt}{2.6ex}}       

\begin{document}

\title[Large-deviation statistics with primordial non-Gaussianity]{Hunting high and low: Disentangling primordial and late-time non-Gaussianity with cosmic densities in spheres}
\author[ C.~ Uhlemann, E. Pajer, C. Pichon, T. Nishimichi, S. Codis and  F.~Bernardeau]{
\parbox[t]{\textwidth}
{C. Uhlemann$^{1}$, E. Pajer$^1$, C. Pichon$^{2,3}$, T. Nishimichi$^{4,5}$, S. Codis$^{6,2}$, F.~Bernardeau$^{2,7}$}
\vspace*{6pt}\\
$^{1}$  Institute for Theoretical Physics \& Center for Extreme Matter and Emergent Phenomena,
 Utrecht University, Princetonplein 5, 3584CC Utrecht, Netherlands\\
$^{2}$ Sorbonne Universit\'es, UPMC Univ Paris 6 et CNRS, UMR 7095,
 Institut d'Astrophysique de Paris, 98 bis bd Arago, 75014 Paris, France \\
$^{3}$ School of Physics, Korea Institute for Advanced Study (KIAS), 85 Hoegiro, Dongdaemun-gu, Seoul, 02455, Republic of Korea\\
$^{4}$ Kavli Institute for the Physics and Mathematics of the Universe (WPI), 
 The University of Tokyo, 5-1-5 Kashiwanoha, Kashiwa 277-85 83, Japan\\
$^{5}$ {CREST, JST, 4-1-8 Honcho, Kawaguchi, Saitama, 332-0012, Japan}\\
$^{6}$ Canadian Institute for Theoretical Astrophysics, University of Toronto, 60 St. George Street, Toronto, ON M5S 3H8, Canada\\
$^{7}$ {CNRS \& CEA, UMR 3681, Institut de Physique Th\'eorique, F-91191 Gif-sur-Yvette, France}
}
\maketitle
\begin{abstract}
Non-Gaussianities of dynamical origin are disentangled from primordial ones using the formalism of large deviation statistics with spherical collapse dynamics. This is achieved by relying on accurate analytical predictions for the one-point probability distribution function (PDF) and the two-point clustering of spherically-averaged cosmic densities (sphere bias). Sphere bias extends the idea of halo bias to intermediate density environments and voids as underdense regions. In the presence of primordial non-Gaussianity,  sphere bias displays a strong scale dependence relevant for both high and low density regions, which is predicted analytically.
The statistics of densities in spheres are built to model primordial non-Gaussianity via an initial skewness with a scale-dependence that depends on the bispectrum  of the underlying model. The analytical formulas with the measured nonlinear dark matter variance as input are successfully tested against numerical simulations. For local non-Gaussianity with a range from $f_{\rm NL}=-100$ to $+100$ they are found to agree within 2\% or better for densities $\rho\in[0.5,3]$ in spheres of radius 15 Mpc$/h$ down to $z=0.35$.
The validity of the large deviation statistics formalism is thereby established for all observationally relevant local-type departures from perfectly Gaussian initial conditions. The corresponding estimators for the amplitude of the nonlinear variance $\sigma_8$ and primordial skewness $f_{\rm NL}$ are validated using a fiducial  joint maximum likelihood experiment.
The influence of observational effects and the prospects for a future detection of primordial non-Gaussianity from joint one- and two-point densities-in-spheres statistics are discussed.
 \end{abstract}

\section{Introduction}

Deviations from primordial Gaussianity are key to understand the physics of  the early universe, in particular inflation, and will soon be tested via a number of ambitious, deep, wide-field Large Scale Structure (LSS) surveys like Euclid \citep{Euclid}, DES \citep{DES}, DESI \citep{DESI}, LSST \citep{LSST} and SPHEREx \citep{SPHEREx}. To pin down the parameters that characterize primordial non-Gaussianity (pNG) as accurately as possible, it is wise to consider the information contained within different probes of LSS in light of our ability to disentangle primordial from nonlinear effects caused by  gravitational evolution. 

This paper revisits the question of  how much information about pNG can be extracted from simple one- and two-point densities-in-spheres measurements within the large scale structure of matter, hence without shape information from the three-point correlation function or the bispectrum.
More precisely, it will consider the full one-point PDF of dark matter densities in spheres thereby consistently including cumulants of all orders. It updates and generalises the analyses based on perturbative predictions for skewness in \cite{FryScherrer94,Durrer00} and kurtosis in \cite{Chodorowski96}, spherical-collapse inspired predictions for cumulants in \cite{Gaztanaga98} or the full PDF derived from a steepest-descent approach in \cite{Valageas02nonG}, as well as the purely simulation-based approach to a joint study of variance, skewness and kurtosis in \cite{Mao14}. It  will relate the statistics of densities-in-spheres \citep[a refined wording for densities-in-cells emphasising the symmetry of the shape of the cells, see e.g.][]{Bernardeau2014} to previous studies of PDFs of the dark matter density \citep{Grossi08} and related works on the impact of primordial non-Gaussianity on abundances of halos (as well as galaxies and clusters residing therein) from Press-Schechter or excursion set based theories \citep{Matarrese00,Scoccimarro04,LoVerde08,Jimenez09,LamSheth09,LamShethDesjacques09,dAmicoHalo11} and voids as well as galaxies residing therein \citep{Kamionkowski09,Song09,dAmicoVoid11}. 

Extending  one-point statistics estimators, this paper also determines  the two-point sphere bias describing the clustering of densities in spheres in the presence of pNG. 
Building upon previous works on scale-dependent halo bias in real and redshift space \citep[in particular][]{Dalal08,MatarreseVerde08,Desjacques09,Valageas10}, it generalises the underlying idea to the two-point bias of average densities in spheres that simultaneously probes intermediate and extreme density environments such as voids and halos.

The outline of the paper is the following: Section~\ref{sec:context} puts densities-in-spheres statistics in the context of current efforts to constrain pNG using LSS. Section~\ref{sec:pNG} briefly reviews the pillars of primordial non-Gaussianity and describes how primordial non-Gaussianity affects the cumulants of the smoothed density field by the example of local non-Gaussianity.  Section~\ref{sec:LDP}  recalls the large deviation principle that allows to obtain one- and two-point statistics for the density-in-spheres. The relationship between the initial density PDF, the initial rate function and correlations at large separation is in particular extended  to non-Gaussian initial conditions.  The spherical collapse model and the saddle-point approximation applied to a log-transformed density are implemented in order to obtain the final non-Gaussian one-point PDF and two-point bias of densities-in-spheres.  Section~\ref{sec:validation}  presents the analytical predictions for the one-point density PDF and two-point sphere bias and compares them to measurements from an $N$-body simulations performed with non-Gaussian initial conditions. Section~\ref{sec:ML}  discusses how the analytical PDFs can be used to constrain primordial non-Gaussianity based on densities-in-spheres observables that are accessible from large scale structure  and addresses promising perspectives. Section~\ref{sec:conclusion} wraps up. Appendix~\ref{app:details} presents generalisations and details of the treatment in the main text.

\section{Context: pNG  in all shapes and sizes}
\label{sec:context}
For the purpose of constraining pNG, two aspects determine how advantageous a particular observable is. First, how much information about pNG the observable contains, and second how well theoretical errors and observational biases can be controlled in order to extract this information and disentangle primordial from evolution effects. The following provides a very short overview of the different probes used to constrain primordial non-Gaussianity using the CMB and LSS and motivates why densities-in-spheres statistics provides complementary information that can prove useful for constraining pNG.

For an in-depth discussion of inflationary models that generate pNG, relevant references and further details  the reviews in \cite{Bartolo04Review,Chen10Review,DesjacquesSeljak10Review,Verde10,Takahashi14Review,CITAworkshop14,Renaux-Petel15Review}
are of interest.

\subsection{CMB vs LSS}
Currently, the best constraints on pNG come from the CMB, hence essentially  from  two-dimensional information at one redshift; it is however a rather direct and clean probe of density fluctuations, which only underwent linear evolution.
 Combining temperature and polarization in the latest Planck results \citep{PlanckPNG15} gives 1-sigma intervals for the amplitude of primordial skewness depending on the shape of the primordial bispectrum of $f_{\rm NL}^{\rm loc}= 0.8 \pm 5.0$, $f_{\rm NL}^{\rm equil}= -4 \pm 43$ and $f_{\rm NL}^{\rm ortho}= -26 \pm 21$. 
Following the standardised normalisation method for $f_{\rm NL}$ from \cite{FergussonShellard09}, observational limits and errors on $f_{\rm NL}^{\rm loc}$ can be consistently compared for different models, finding that despite appearances, upon this normalization current bounds from the CMB are of comparable order for the different shapes. 

Reaching the theoretically interesting benchmark of $f_{\rm NL}\lesssim  1$ will require to extract yet more information that is in principle available within large scale structure (LSS) because it probes three-dimensional information for many modes over a wide range of redshifts (that can be obtained accurately from spectroscopic surveys). Unfortunately, this information is obscured by nonlinear gravitational evolution, galaxy bias and redshift space distortions and hence requires accurate theoretical modelling to disentangle dynamically induced from primordial non-Gaussianity. Note that, most LSS-based probes are only sensitive to a particular range of bispectrum shapes and hence their ability to constrain the different parameters varies. Eventually one should of course  consider combined constraints from the cosmic microwave background and large-scale structure  \citep[see e.g.][]{Giannantonio14a,Giannantonio14b}, especially to constrain scale-dependent non-Gaussianity, see e.g. \cite{LoVerde08,Sefusatti09,Schmidt10,Becker12}.
For the next generation `Stage-4' ground-based CMB experiment, CMB-S4 \citep{CMB-S4}, the expected factor of improvement over Planck is slightly more than a factor of 2 which will be insufficient to reach the interesting theoretical benchmark for local $f_{\rm NL}$.
 
\subsection{Abundance and large scale clustering of rare objects}
Both the abundance and the clustering signal of rare objects, namely extrema of the density distribution such as halos or voids, are suitable to probe  pNG. The effect of pNG on the halo mass function and cluster number counts has been studied theoretically in \cite{Matarrese00,LoVerde08,AfshordiTolley08,Valageas10,Musso12} within the framework of an extended Press-Schechter theory and compared to simulations in \cite{Grossi08}. \cite{Verde13} considered multi-variate PDFs that characterize the effect of $f_{\rm NL}$ on different observables that can be obtained from CMB maps, but also the PDF of the linearly extrapolated and smoothed density field. In all those studies it has been found that the presence of pNG modifies the abundances of halos and voids increasingly with the rarity of the object.  While in principle the presence of pNG might also alter the density profiles of rare objects, the impact on halo profiles has been observed to be small \citep{MoradinezhadDizgah13} and a similar study for voids that is underway shows no perceptible differences on void profiles.\footnote{private communication with N. Hamaus, \& K. Chan}

Furthermore, following an observation from \cite{Dalal08,Desjacques09}, pNG leads to strongly scale-dependent bias that impacts the large-scale clustering of rare objects (peaks of the density distribution such as halos) which hence can be used to constrain pNG from large scale structure \citep{Slosar08}. Subsequently, pNG-induced halo bias and its dependence on scale, mass and redshift in the limit of high peaks and large separations has been theoretically investigated in \cite{MatarreseVerde08}, compared to simulations in \cite{Grossi09} and refined in a treatment in real space in \cite{Valageas10}. 

Scale-dependent bias has already been used to put constraints on pNG based on galaxies and quasars within the SDSS in \cite{Slosar08,Ross13,Leistedt14} finally reaching an individual constraint on $-49<f_{\rm NL}<31$ (95\% confidence) for local pNG. The prospects of constraining pNG in future surveys with two-point statistics like the 3D galaxy power spectrum in redshift space, the angular galaxy power spectrum and the projected weak-lensing shear power spectrum have been investigated in
\cite{Giannantonio12} finding a forecasted marginalized error of $\sigma(f_{\rm NL}) \simeq 3$ on local pNG for
Euclid-like survey combining all probes. This is in line with current forecasts of $\sigma(f_{\rm NL})\simeq 4-5$ (competitive with Planck) for local pNG achievable with Euclid \citep{Euclid16} by combining galaxy clustering from spectroscopic and photometric data as well as weak lensing.

With recent advancements it should be possible to improve constraints for local pNG from the two-point clustering of single populations of large-scale structure tracers, which are already at the level of pre-Planck constraints from the CMB. To this end, \cite{Seljak09} demonstrated how the effect of cosmic variance  for the estimation of pNG can be mitigated using two sets of tracers which, at large scales, are deterministically biased with respect to dark matter. This allows to extract the scale-dependent relative bias due to pNG with an accuracy set by the noise due to the discrete (Poisson) sampling of the density field rather than cosmic variance. \cite{Hamaus11} investigated constraints on local pNG from two-point statistics when using two optimization strategies -- avoiding sampling variance by comparing multiple tracers of different bias and suppressing shot noise by optimally weighting halos of different mass.

\cite{ByunBean15} investigated the dependence of the halo mass function and scale-dependent halo bias on the primordial bispectrum shape. They found that the scale-dependent bias on large scales probes general squeezed configurations from the primordial bispectrum, while the scale-dependent bias on smaller scales and the halo mass function are more sensitive to a broader range of shapes. Looking at the not-quite-large scales is also interesting because effects of General Relativity (GR) become significant in the very large-scale regime, precisely where the scale-dependent bias induced by pNG is strongest \citep[see e.g.][]{Bruni12,Camera15}, although it has been shown that GR effects on scale-dependent bias are not degenerate with $f_{\rm NL}$ \citep{Yoo12}.

\subsection{Bi- and trispectrum of the galaxy distribution}

Since constraining pNG aims at probing the component of the initial bispectrum induced by non-Gaussianity, looking at its observable late-time equivalent, the galaxy bispectrum, is a natural possibility that in particular provides means of distinguishing between different shapes. For local pNG with small amplitude $f_{\rm NL}$, one can show that the primordial bispectrum estimator is optimal and equivalent to calculating the full likelihood of the data \citep{Creminelli07}.
Unfortunately, the theoretical modelling of the late-time dark matter bispectrum induced by gravitational evolution already proves to be difficult. In addition to the modelling of the dark matter bispectrum, observational effects such as tracer bias \citep{Kaiser84,dekel87} and redshift space distortions \citep{Kaiser87,tns} need to be included in the analysis. 

Constraints on $f_{\rm NL}$ from measurements of the galaxy bispectrum in redshift surveys have been considered as early as in \cite{Scoccimarro04} concluding that planned galaxy surveys at high redshifts can in principle give pNG constraints comparable to, or even better than, those from CMB experiments. This was confirmed in \cite{SefusattiKomatsu07} by performing a Fisher matrix analysis for the galaxy bispectrum at high redshift with pNG and nonlinear but local bias, including a study of how constraints of pNG improve with volume, redshift range, as well as the number density of galaxies. These works relying on tree-level perturbation theory for the matter bispectrum in the presence of pNG have been subsequently extended to include one-loop corrections \citep[see e.g.][]{JeongKomatsu09,Sefusatti09pt,Bernardeau10,Matsubara11} that improve the agreement with numerical simulations \citep{Sefusatti10}. 
Following up on observations of scale-dependent bias in the two-point clustering of peaks, \cite{JeongKomatsu09} considered the bispectrum of galaxies as peaks finding that the effect of local $f_{\rm NL}$ on the galaxy bispectrum cannot be obtained by replacing the linear bias in the galaxy bispectrum with the scale- dependent bias obtained for the power spectrum.
Those perturbative analyses based on tree-level perturbation theory and a local bias model have been generalised in \cite{Baldauf11} to incorporate a multivariate bias expansion \citep{Giannantonio10} and the peak-background split method \citep{Slosar08}.
The validity of this complete tree-level approximation at large scales has been established by simulations with local pNG \citep{Sefusatti12} and used to get a rough Fisher-matrix based estimate for an expected improvement of a few over the halo power spectrum for a combined  power- and bispectrum analysis.

A more recent comparison of different perturbative and phenomenological models for the matter bispectrum with Gaussian and non-Gaussian initial conditions against numerical simulations has been performed in \cite{Lazanu16,Lazanu17}, see also references therein. It has been found that among the perturbative approaches, the Effective Field Theory of Large Scale Structure \citep[EFTofLSS as introduced in][]{Baumann12} extends the range of validity furthest on intermediate scales. However, this comes at the cost of introducing free extra parameters requiring calibration on simulations or marginalisation. The EFTofLSS includes those free parameters to encode our ignorance of the small scale physics that cannot be captured with perturbation theory \citep{Bernardeau02} and hence to trace theoretical errors. 

Based on the matter bispectrum from EFTofLSS including pNG \citep{Assassi15}, estimates for  the amount of primordial non-Gaussianity while including theoretical errors in the modelling of the bispectrum were obtained in \cite{Baldauf16EFT,Welling16}. They found that accounting for these theoretical errors can weaken constraints considerably, for example degrading by a factor 5 from  the idealised forecast for Euclid of $\sigma(f_{\rm NL})\simeq 0.45$ (with floating bias)  in \cite{Tellarini16} down to $\sigma(f_{\rm NL})\simeq 1.8$ (with priors on EFT parameters)  in \cite{Welling16}. In the latter it has also been established that assuming the wrong shape for the theoretical error might lead to a false detection of pNG, highlighting how difficult the theoretical modelling proves to be. In addition, recent studies show that relativistic effects are relevant for the bispectrum at an order $\sigma(f_{\rm NL})\simeq 1$ and hint at a degeneracy with local pNG \citep{DiDio17}.

Regarding higher order galaxy statistics, \cite{VerdeHeavens01} argued, based on an idealised case with essentially linear evolution, that a measurement of the trispectrum, which has weaker dependence on nonlinear clustering, may provide pNG constraints complementary to the bispectrum in the future. 

\subsection{Densities-in-spheres statistics}

The statistics of densities-in-spheres (or counts-in-cells as their counterpart for discrete tracers) 
entail in particular the one-point PDF of finding a certain average density in a top-hat sphere of given radius and the two-point clustering of those spheres given their average density. While those statistics are related to the distribution of tracers and rare objects 
such as halos and voids, they probe a different regime than the bispectrum of tracers or the abundance and clustering of rare objects and hence can improve our ability to probe pNG.
One-point densities-in-spheres statistics already contains {\it some} of the information that is in all higher-order correlation functions  of the smoothed density field, namely their zero separation limit -- characterizing the cumulants and thereby the shape of the PDF. The cumulants of the smoothed density field can be predicted accurately using perturbation theory or spherical collapse starting from Gaussian statistics \citep{Bernardeau94smoothing}, but can also be extended to include pNG in an expansion around Gaussian initial conditions \citep{FryScherrer94,Chodorowski96,Gaztanaga98}. 
An alternative approach to the statistics of densities-in-spheres is the excursion set model \citep{Sheth98}, which can be used to relate the halo mass function to the dark matter distribution in Eulerian space. Recently, also excursion set theory has been extended to non-Gaussian initial conditions \citep{MussoSheth14}. Since the PDF includes cumulants of all orders, it comprises more information than (a finite set of) single cumulants which is especially relevant in the tails that are generated through nonlinear gravitational dynamics \citep{Bernardeau94rareevents} , but also sensitive to pNG. The full density PDF and its cumulants have been considered for models with weak pNG arising from standard slow-roll inflation \citep{Gaztanaga98,Valageas02nonG} and strong pNG arising from dimensional scaling models \citep{Turok91,Gaztanaga96,White99,Scoccimarro00}. 
Densities-in-spheres statistics include some higher-order information beyond the bispectrum, but they only incorporate a very specific part of its rich shape-dependent information. This may seem to be a significant drawback when considering  statistical information content for constraining pNG, but it is mitigated by  our ability to build robust estimators through  ab initio theoretical modelling. While cumulants of average densities in spheres can be predicted robustly from simple spherical collapse dynamics for both Gaussian \citep{Bernardeau94rareevents} and close-to-Gaussian initial conditions \citep{Gaztanaga98,Valageas02nonG}, modelling the bispectrum requires significantly more complex perturbative methods as discussed in the previous paragraph.

The one point PDF includes information about abundances of rare objects (halos and voids) simultaneously in both tails, hence offering a unified treatment of all density environments. Similarly, two-point densities-in-spheres statistics, that can be computed for large separations \citep{Bernardeau96bias,AbbasSheth07,Codis16correlations,Uhlemann17Kaiser} for Gaussian initial conditions, captures the large-scale clustering of rare objects (extreme density environments occurring in halos or voids) in the tails but also encodes differences in clustering for more common density environments. Given that the most competitive pNG constraints from LSS to date come from the scale-dependence of halo bias, it is promising to generalise this idea to a more diverse range of density environments. The unified treatment of both high and low densities could in particular help to disentangle otherwise degenerate effects such as $f_{\rm NL}$ and $g_{\rm NL}$.

In spirit related to the PDF of densities in spheres is the pairwise velocity PDF whose sensitivity on local pNG has been investigated in \cite{LamNishimichi11} where 5-10\% effects of primordial non-Gaussianity on the PDF were found for $f_{\rm NL}=\pm 100$, but its accurate theoretical modelling proved to be difficult.

\section{Primordial non-Gaussianity \& cumulants}
\label{sec:pNG}
Let us first express the leading order cumulants in terms of local pNG parametrized by $f_{\rm NL}$. We refer the readers to  Appendix~\ref{app:beyondlocal} for next-to-leading order terms and non local expansions.  

\subsection{Local primordial non-Gaussianity }
For simplicity,  the main text of the paper  considers leading order, local (quadratic) primordial non-Gaussianity where the non-Gaussian field $\Phi_{\rm NG}$ (the gravitational potential) is expressed in terms of a Gaussian field $\Phi_G$
\begin{equation}
\label{eq:lpnG}
\Phi_{\rm NG}= \Phi_G+ f_{\rm NL} \left(\Phi_G^2-\langle\Phi_G^2\rangle\right) \,,
\end{equation}  
with constant parameter $f_{\rm NL}$. For a generalisation including the cubic order with $g_{\rm NL}$, see Appendix~\ref{app:NLOfnl}. In principle, one can express the non-Gaussian density in terms of the Gaussian density and the gravitational potential using the identity
\begin{equation}
\label{eq:lpnGlaplacian}
\Delta \Phi_{\rm NG} = \Delta \Phi_{\rm G} + 2f_{\rm NL} \left(\Phi_{\rm G}\Delta\Phi_{\rm G} + |\nabla \Phi_{\rm G}|^2\right) \,,
\end{equation}
where $\Delta$ is the Laplacian and hence the left hand side is proportional to the density. In practice, it is often more convenient to work in Fourier space\footnote{ The Fourier convention $\delta(\vx)= (2\pi)^{-3} \int \dd^3 k \exp(i\vk\cdot\vr)\hat \delta(\vk)$ is used throughout. For the sake of simplicity, we will omit the hat on the Fourier transforms.}
 where the Poisson equation becomes a simple multiplication and smoothing  operates  by multiplication rather than convolution.

Let us quantify how primordial non-Gaussianity influences the PDF of initial densities which are smoothed over a certain radius. Hence, one needs to relate the linearly evolved density field to the primordial curvature perturbation $\zeta$ encoding the information of non-linearities produced during and after inflation. We assume the dimensionless primordial power-spectrum of the (comoving) curvature perturbation to be of the form
\begin{equation}
\label{eq:primPS}
\Delta_\zeta(k)=A_s(k_0) (k/k_0)^{(n_s-1)}\,,
\end{equation}
with the amplitude of the scalar power spectrum $A_s\simeq 2.5 \times 10^{-9}$ measured at the pivot scale, $k_0=0.002/h$ Mpc$^{-1}$, and the primordial spectral index $n_s\simeq 0.96$ \citep[consistent with recent limits from Planck][]{Planck16inflation}, encoding initial conditions with a small departure from scale-invariance as predicted from slow-roll models of inflation \citep{Mukhanov81}. 
The gravitational potential $\Phi$\footnote{Note that in conformal Newtonian gauge, the gravitational potential is related to the Bardeen potential as $\Phi_B=-\Phi$.} is related to the curvature perturbation $\zeta$, which is preserved on super-Hubble scales, by a constant factor of $3/5$ during the matter domination era such that its power spectrum is obtained as
\begin{align}
P_\Phi(k)=\left(\frac{3}{5}\right)^2 \frac{2\pi^2}{k^3} \Delta_\zeta(k)\,.
\end{align}
The gravitational potential $\Phi$ and the linear perturbation to the matter density $\delta_p$ at redshift $z$ are related through the Poisson equation on sub-Hubble scales 
\begin{equation}
\delta_p(k,z)=g(k)T(k) D(z)\Phi_{\rm NG}(k) \ , \ g(k) = \frac{-2}{3\Omega_m}\left(\frac{k}{H_0}\right)^2\,,
\end{equation}
where $T(k)$ is  the transfer function of perturbation (which goes to unity on very large scales $T(k)\stackrel{k\rightarrow 0}{\longrightarrow}1$) and encodes the suppression of power for modes that entered the horizon before matter-radiation equality, $D(z)$ is the linear growth factor which is related to the redshift as $D(z) = (1+z)^{-1}$ in matter domination, $\Omega_m$ is the present time fractional matter density and $H_0 = 100h \text{km s}^{-1}\text{Mpc}^{-1}$ the Hubble constant.
Finally,  the spherical top-hat filtering on the density field can be applied to obtain 
\begin{equation}
\delta_R(k) =W_{\rm 3D}(kR) \delta_{p}(k)\,,
\end{equation}
where $W_{\rm 3D}$ is the Fourier transform of  the top-hat filter
\begin{equation}
W_{\rm 3D}(x)=\frac{3}{x^{2}}(\sin(x)/x-\cos(x))\,.
\end{equation}
Now, primordial non-Gaussianity for the gravitational potential can be expressed conveniently as
\begin{align}
\delta_{R,\rm NG}(k,z)&= \alpha_R(k,z) \Phi_{\rm NG}(k)\,,\\
\notag \alpha_R(k,z)&= W_{\rm 3D}(kR) g(k)T(k) D(z) \,.
\end{align}
The key assumption that allows for a general treatment of a mildly non-Gaussian field is that it can be expanded as a local functional of an underlying Gaussian field. On the other hand, any form of nonlocality that can be expressed as a convolution in real space can also be handled in a similar way by suitably modifying our definition of the kernel functions $\alpha_R$. This means that our results can be easily extended to a primordial bispectrum of an arbitrary shape. 

\subsection{Leading order cumulants of the smoothed density field}

For one-point statistics, the skewness of the smoothed density field is the leading order correction to Gaussian initial conditions. To this order, the effect of local primordial non-Gaussianity on the skewness is  given by \citep[see also][]{Matarrese00}
\begin{align}
\label{eq:skewnesslocal}
\kappa_{3}(R)&\equiv \langle \delta_{R,\rm NG}^3\rangle\,,\\
\notag &\simeq \frac{3 f_{\rm NL}}{4\pi^4} \! \int \! \dd k_1\, k_1^2\, \alpha_R(k_1) P_\Phi(k_1) \! \int \! \dd k_2 \,k_2^2\, \alpha_R(k_2) P_\Phi(k_2)\\
 & \qquad \times \int_{-1}^1\dd\mu_{12} \, \alpha_R\left(\sqrt{k_1^2+k_2^2 + 2k_1k_2\mu_{12}}\right)\,.
\end{align}
For the non-local cases, the skewness (and also higher order cumulants) of the smoothed density field can be obtained from the primordial bispectrum $B_\Phi$ of arbitrary shape, as discussed in Appendix~\ref{app:beyondlocal} based on equation~\eqref{eq:skewnessbispectrum}. Because of the smoothing involved, the scale-dependence for different shapes is typically very similar, as demonstrated in Figure~\ref{fig:scaledepskewnessmodels}, so the focus is here on local non-Gaussianity for simplicity. 

It is often convenient to consider cumulants which are rescaled by certain powers of the variance, such as the reduced cumulants $\tilde S_n$ \citep[which remain approximately constant through late time gravitational evolution for Gaussian initial conditions][]{Colombi97} or the rescaled cumulants $\epsilon_n$ \citep[which are approximately independent of smoothing scale and robust against linear growth as it cancels in the ratio][]{dAmicoHalo11}. They enter any cumulant-based expansion -- e.g the Edgeworth expansion -- of the cosmic PDFs around Gaussian kernels
\begin{align}
\label{eq:redcumulants}
\tilde S_n = \frac{\kappa_n}{\sigma^{2(n-1)}} \,,\quad \varepsilon_n = \frac{\kappa_n}{\sigma^{n}}=\sigma^{n-2}\tilde S_{n} \,.
\end{align}
 The tilde is used to distinguish the primordial reduced cumulants from the ones that are gravitationally evolved. The rescaled cumulants $\epsilon_n$ are to a very good approximation perturbatively ordered, as pointed out in \cite{dAmicoHalo11} (where however a top-hat filter in $k$-space was used instead of a spherical top-hat filter in real space as considered here). Note that the smallness parameter is $\varepsilon_3\propto f_{\rm NL} A^{1/2}$ where $A_s \simeq 2.5 10^{-9}$ is the amplitude of the primordial power-spectrum $P_\Phi$, similarly  $\varepsilon_4\propto \varepsilon_3^2\propto f_{\rm NL}^2 A$ or  $\varepsilon_4\propto g_{\rm NL} A$. Hence, for reasonably small $f_{\rm NL}\simeq 100$ one can typically neglect next-to-leading order contributions appearing in form of a kurtosis and a second-order correction to the variance.  Appendix~\ref{app:NLOfnl} presents the next-to-leading order results (including the kurtosis and a second-order correction to the variance) for a local model including both $f_{\rm NL}^2$ and $g_{\rm NL}$.

For the two-point statistics,  the leading order mixed cumulant $\kappa_{12}(r)$ is also needed; it is obtained as 
\begin{align}
\label{eq:mixedskewness}
\kappa_{12}(r) &= \langle \delta_{R,\rm NG}(\vx+\vr)\delta_{R,\rm NG}^2(\vx)\rangle \\
&= \frac{2f_{\rm NL}}{(2\pi)^6} \int\!\!\!\! \int \!  \dd^3k_1 \dd^3k_2 \alpha_R(k_1)\alpha_R(k_2)\alpha_R(k) \\
\notag &\qquad \qquad \times  P_\Phi(k_1)\left[P_\Phi(k_2)+2P_\Phi(k)\right] \exp[i \vk\cdot \vr]\,,
\end{align}
where $\vr$ is the separation between the two points and $\vk=\vk_1+\vk_2$. Integrating over the angle between $\vk$ and $\vr$ yields
\begin{align}
\notag
\kappa_{12}(r) 
=\frac{f_{\rm NL}}{4\pi^4} & \int\!\! \dd k\, k^2 \alpha_R(k)\, \text{sinc}(kr) \int\!\! \dd k_1\, k_1^2 \alpha_R(k_1) P_\Phi(k_1)\\
 &\ \int_{-1}^1\dd\mu_k\, \alpha_R(k_2)  \left[P_\Phi(k_2)+2P_\Phi(k)\right] \,,
 \label{eq:3ptcorr}
\end{align}
where $k_2=|\vk-\vk_1|=\sqrt{k_1^2+k^2-2k_1k\mu_k}$ and $\mu_k$ is the cosine of the angle between $\vk$ and $\vk_1$. For a generalisation of the formula for arbitrary bispectra see Appendix~\ref{app:beyondlocal}, in particular equation~\eqref{eq:mixedskewnessbispectrum}. Naturally, when  evaluating the leading order mixed cumulant at zero separation, one correctly recovers the skewness from equation~\eqref{eq:skewnesslocal}.

\section{Constructions of the density PDFs}
\label{sec:LDP}

Let us first briefly present the large deviation principle that allows us to obtain one-point statistics for the density in a sphere of fixed final radius,
focusing on the extension to non-Gaussian initial conditions.  Readers already familiar with the formalism for Gaussian initial conditions may skip to the extension to primordial non-Gaussianity given in Section~\ref{subsec:pNGratefunction} and then proceed with the phenomenological effects in Section~\ref{subsec:phenopNG}.

\subsection{Large deviation statistics in spheres}
\label{sec:SC}

When considering a highly symmetric observable such as the density in spheres, one can argue that the most likely dynamics (amongst all possible mappings between the initial and final density field) is the one respecting the symmetry \citep{Valageas02}.\footnote{{This is a result of the so-called contraction principle in the context of large deviation theory as explained in \cite{LDPinLSS},
which formalizes the idea that amongst all unlikely fates (in the tail of the PDF) the least unlikely one (i.e. the spherical collapse solution) dominates.}}  Spherical symmetry allows us to take advantage of the fact that non-linear solutions to the gravitational dynamics are known explicitly in terms of the spherical collapse model.

Let us denote $\rho_{\rm SC}(\tau)$ the non-linear transform of an initial fluctuation with linear density contrast, $\tau$, in a sphere of radius $R_{\rm ini}$, to the final density $\rho$ (in units of the average density) in a sphere of radius $R$ according to the spherical collapse model
\begin{equation}
\rho=\rho_{\rm SC}(\tau)\,, 
\quad {\rm with}
\quad
\rho R^{3}= R_{\rm ini}^{3}\,, 
\label{eq:rho2tau}
\end{equation}
where the initial and final radii are connected through mass conservation. An explicit possible fit for $\rho_{\rm SC}(\tau)$ is given by
\begin{equation}
\rho_{\rm SC}(\tau)=(1-\tau/\nu)^{-\nu} \ \Leftrightarrow \ \tau_{\rm SC}(\rho)= \nu (1-\rho^{-1/\nu})\,, \label{eq:spherical-collapse}
\end{equation}
where $\nu$ can be adjusted to the actual values of the cosmological parameters ($\nu=21/13$ provides 
a good description of the spherical dynamics for an Einstein-de Sitter background for the range of $\tau$
values of interest).

Thanks to this analytic spherical collapse model, the one-point PDF and bias functions of cosmic densities in concentric spheres, brought about by  non-linear gravitational evolution, can be predicted explicitly from the given (close-to Gaussian) initial conditions. 

\subsection{Initial one-point PDF and decay-rate function}
The large-deviation principle yields a formula for the PDF of finding a certain density in a sphere of a given radius given the statistics of the initial conditions. The initial decay-rate function $\Psi^{\rm ini}_{R_{\rm ini}}$ encodes the exponential decay of the PDF given the cumulants of the initial field. While in general the initial PDF is related to the cumulant generating function via an inverse Laplace transform, this relation is well approximated by a saddle-point approximation for close-to-Gaussian initial conditions. Then, the initial PDF can be written as
\begin{align}
 \label{eq:saddlePDFlog-ini}
\mP_{R_{\rm ini}}^{\rm{ini}}(\tau)&= \sqrt{\left|\frac{\partial^{2}\Psi^{\rm ini}_{R_{\rm ini}}}{\partial \tau^2}\right|} \frac{ \exp\left[-\Psi^{\rm ini}_{R_{\rm ini}}(\tau)\right]}{ (2 \pi)^{1/2}} 
\,.
 \end{align}
 
 \subsubsection{Gaussian initial conditions}

For Gaussian initial conditions, the  initial decay-rate function is simply given by a quadratic form in the initial density contrast $\tau$ in a sphere of radius $R_{\rm ini}$ where the linear variance $\sigma^{2}_\lin$ encodes all dependency with respect to the initial power spectrum for the Gaussian field
\begin{align}
\label{eq:PsiDefIni}
\Psi^{\rm ini}_{R_{\rm ini}}(\tau)&=\frac{1}{2}\frac{\tau^2}{\sigma^2(R_{\rm ini})}\,, \\
\label{eq:variance}
\sigma^{2}_\lin(R_{\rm ini}) &= \! \int \! \! \frac{\dd k}{2\pi^2} \,  k^2\, P_{\lin }(k)W_{\rm 3D}^2(kR_{\rm ini})\,.
\end{align}
Note that in this case equation~\eqref{eq:saddlePDFlog-ini} is merely an  unusual rewriting of a Gaussian distribution, emphasizing the central role of the rate function~\eqref{eq:PsiDefIni}. The next step involves obtaining a suitable form of the initial decay-rate function for non-Gaussian initial conditions based on a cumulant expansion whose result is given in equation~\eqref{eq:ratefctpNGsaddle} below. For this analysis, the key ingredient of the decay-rate function is the scale-dependent skewness of the smoothed density field which can be computed for a local model using equation~\eqref{eq:skewnesslocal} and for any given bispectrum shape according to equation~\eqref{eq:skewnessbispectrum}.

\subsubsection{Non-Gaussian initial conditions: cumulant expansions}
\label{subsec:pNGratefunction}
The complete series of cumulants of a PDF is related to the cumulant generating function whose successive derivatives give back the cumulants.
Taking the modifications in the cumulants of the smoothed density up to quadratic order in non-Gaussianity (including skewness, kurtosis and second-order variance) into account leads to the following truncated cumulant generating function
\begin{align}
\label{eq:cumgen}
\tilde\varphi_{R_{\rm ini}}(\lambda) &= \frac{\lambda^2}{2}\left[\kappa_{2,{R_{\rm ini}}}^{(0)+(2)}+\frac{\lambda}{3}\kappa_{3,{R_{\rm ini}}} + \frac{\lambda^2}{12} \kappa_{4,{R_{\rm ini}}}\right]\,,
\end{align}
where $(0)$ and $(2)$ refer to the zeroth and next-to-leading order in $f_{\rm NL}$ respectively. 
Note that this equation only holds for local quadratic pNG as is assumed throughout the main text.
Performing a Legendre transformation of the cumulant generating function then gives the initial rate function 
\begin{align}
\label{eq:ratefctpNGsaddle}
\Psi_{R_{\rm ini}}(\tau)
 &\simeq \frac{\tau^2}{2\sigma(R_{\rm ini})^2} \left[1-\frac{\varepsilon_3(R_{\rm ini})}{3}\frac{\tau}{\sigma(R_{\rm ini})} - \varepsilon_2^{(2)}(R_{\rm ini}) \right. \notag\\
 &\quad \left.+ \left(\frac{\varepsilon_3(R_{\rm ini})^2}{4}-\frac{\varepsilon_4(R_{\rm ini})}{12}\right)\frac{\tau^2}{\sigma(R_{\rm ini})^2} \right]\,,
\end{align}
using equation~\eqref{eq:skewnesslocal} to express the leading order term $\varepsilon_3$, and equations~\eqref{eq:kurtosis}~and~\eqref{eq:variancepNG} for the next-to-leading order terms in terms of the underlying  power spectrum and $f_{\rm NL}$;
see also equations~(50) and (58) in \cite{Matarrese00} and  \cite{Valageas10}.
 Since $\varepsilon$ are the dimensionless initial cumulants, the linear growth cancels out in this quantity but does enter in the variance $\sigma$. 
Since the focus is   in the tails   while not relying on perturbation theory,  $\tau/\sigma_r$ need not be small. This is the main difference of our analysis which hence complements the perturbation-theory based analysis for the expansions for the mass function in \cite{dAmicoHalo11}. Hence, the cumulant-expanded form of the rate function in equation~\eqref{eq:ratefctpNGsaddle} is to be preferred over an Edgeworth expansion around a Gaussian PDF that is discussed in Appendix~\ref{app:Edgeworth}. In order for the truncation in the series of primordial cumulants to be accurate, one must have that $1\gg \varepsilon_{3}{\tau}/{\sigma} \gg \varepsilon_{4} \left({\tau}/{\sigma}\right)^2$. This in principle limits how far into the tails one is allowed to probe when just considering primordial skewness. To obtain an estimate for the allowed range given an $f_{\rm NL}$, one can use the spherical collapse solution \eqref{eq:spherical-collapse} to obtain the initial overdensity $\tau\rightarrow\tau(\rho)$ and the radius $R_{\rm ini}=R\rho^{1/3}$ connected to a final density $\rho$ in a sphere of radius R. 
For $f_{\rm NL}\lesssim 100$, higher cumulants are still suppressed by one order of magnitude for typical densities $\rho\in[0.1,10]$ and hence it will be sufficient to account for the primordial skewness for our analysis.

\subsection{One-point PDF for an evolved non-Gaussian field}  
The final decay-rate function is obtained from re-expressing the initial decay-rate \eqref{eq:ratefctpNGsaddle} in terms of the final densities and radii
\begin{equation}
\label{eq:SCreplacement}
\tau_{R_{\rm ini}} = \tau_{\rm SC}(\rho_R)\ , \ R_{\rm ini}=R\rho^{1/3}\,,
\end{equation}
using the spherical collapse mapping equation~\eqref{eq:spherical-collapse}, that corresponds to a saddle-point approximation\footnote{Note that, according to \cite{Valageas10} the effect of a realistic $|f_{\rm NL}|\lesssim 100$ on the density profile of the saddle point is negligible such that also the onset of shell-crossing remains practically unchanged compared to the Gaussian case presented in \cite{Valageas09}.}
and rescaling the linear variance at scale $R$ to its correct nonlinear value
\begin{align}
\label{eq:variancerescaling}
\sigma_{\rm lin}(R_{\rm ini}) \rightarrow \sigma(R_{\rm ini}) \simeq \frac{\sigma(R)}{\sigma_{\rm lin}(R)} \sigma_{\rm lin}(R\rho^{1/3}) \,.
\end{align}
Altogether the decay-rate function reads
\begin{align}
\Psi_{R}(\rho)= \Psi^{\rm ini}_{R_{\rm ini}=R\rho^{1/3}}\left(\frac{\sigma_{\rm lin}(R)}{\sigma(R)}\tau_{\rm SC}(\rho)\right)\,,
\label{eq:PsiDef}
\end{align}
 where $\Psi^{\rm ini}_{R_{\rm ini}}$ is given by equation~\eqref{eq:ratefctpNGsaddle}.
From that decay-rate function, one can obtain the cumulant generating function (via a Legendre transformation) and from there the PDF by performing an inverse Laplace transformation. 
As discussed in \cite{Uhlemann16}, one can use a saddle point approximation to accurately evaluate the corresponding integral and obtain a direct relation between the final decay-rate function and the final PDF. This requires a suitable choice of variable such that the final decay-rate function is convex, i.e. $\Psi''_R>0$ for all densities of interest. One particularly advantageous choice is the logarithmic density $\mu=\log\rho$ for which one can apply the saddle-point approximation to predict the PDF of the logarithmically mapped density and translate this to the PDF of the density field via a simple change of variables
\begin{align}
\notag \mP_R(\rho) &=\mP_{\mu,R}[\mu(\rho)] \left|\frac{\partial\mu}{\partial \rho}\right| = \sqrt{\frac{\partial^{2}\Psi_R}{\partial \mu^2}} \left|\frac{\partial\mu}{\partial \rho}\right| \frac{ \exp\left[-\Psi_R\right]}{ (2 \pi)^{1/2}} \\
&= \sqrt{\frac{\Psi_R''(\rho)+\Psi_R'(\rho)/\rho}{2\pi}} \exp[-\Psi_R(\rho)] \,.
 \label{eq:saddlePDFlog}
 \end{align}
Note that since the decay-rate function $\Psi_R$ appearing here has been written down for the log-density, the variance $\sigma(R)=\sigma_\mu(R)$ that is entering equation~\eqref{eq:PsiDef}  is the variance of the log-density $\mu=\log\rho$.
To ensure a unit mean density after the mapping, one has to shift the log-density appropriately which is equivalent to considering the renormalized density $\tilde\rho=\rho/\langle\rho\rangle$ with the shorthand notation $\langle f(\rho)\rangle= \int_0^\infty \dd\rho\, f(\rho)\mP_R(\rho)$. Furthermore, since the saddle-point method yields only an approximation to the exact PDF, the PDF obtained from equation~\eqref{eq:saddlePDFlog} is not necessarily perfectly normalized (although this effect is usually sub-percent). In practice, one can account for both effects for by considering the normalized PDF
\begin{equation}
\label{eq:saddlePDFrenorm}
\hat\mP_{R}(\rho)=\mP_{R}\left(\rho\frac{\langle\rho\rangle}{\langle 1\rangle}\right)\frac{\langle \rho\rangle}{\langle1\rangle^2}\,.
\end{equation}
Equations~\eqref{eq:cumgen}-\eqref{eq:saddlePDFrenorm} yields the general evolved non Gaussian PDF
of the density field.

\subsection{Initial two-point PDF and sphere bias}
The initial sphere bias used in previous works on densities-in-spheres \citep[see e.g.][]{Bernardeau96bias,Codis16correlations,Uhlemann17Kaiser} is defined as the ratio between the conditional mean density induced by an initial density contrast $\tau$ in a sphere of radius $R_{\rm ini}$ at separation $r$ and the average correlation
\begin{align}
\label{eq:Kaiserbiasini}
b_{\rm ini}(\tau,r)&= \frac{\langle\tau'(\vx+\vr)|\tau(\vx)\rangle}{\langle\tilde\tau'(\vx+\vr)\tilde\tau(\vx)\rangle} = \frac{\displaystyle\! \int \! d\tau' \,\tau'\,\mP_{R_{\rm ini}}(\tau,\tau';r)}{\mP_{R_{\rm ini}}(\tau) \xi_{\rm lin}({R_{\rm ini}},r)} \,,
\end{align}
and can be expressed in terms of the joint PDF of densities in spheres of radius $R_{\rm ini}$ at separation $r$ and their correlation function. The initial correlation function is obtained from the linear power spectrum as
\begin{align}
\label{eq:corrlin}
\xi_{\lin}(R_{\rm ini},r) &=\! \int \! \! \frac{\dd k}{2\pi^2} k^2 P_{\lin}(k)W_{\rm 3D}^2(kR_{\rm ini}) \text{sinc}(kr) \,.
\end{align} Note that this expression only depends on the density $\tau$, the radius of the spheres $R_{\rm ini}$ and their separation $r$ while the densities $\{\tau',\tilde\tau',\tilde \tau\}$ are dummy variables for the evaluation of the spatial expectation values. The sphere bias can be used to express the joint PDF of densities in spheres of radius $R$ at large separation $r\gg R$
\begin{align}
\label{eq:jointPDFlargesep}
\frac{\mP_{R_{\rm ini}}(\tau,\tau';r)}{\mP_{R_{\rm ini}}(\tau)\mP_{R_{\rm ini}}(\tau')}\!\simeq\! 1\!+\!\xi_{\rm lin} (R_{\rm ini},r) b_{\rm ini}(\tau,r)b_{\rm ini}(\tau',r)\,.
\end{align}

\subsubsection{Gaussian initial conditions}
To derive the initial bias, let us follow the procedure described in Appendix C of \cite{Uhlemann17Kaiser}. The covariance matrix of $(\tau,\tau')$  is given by
\begin{equation}
\label{eq:Kaisercovmatrix}
\Sigma_{\rm lin}=\left(
\begin{array}{ccc}
 \sigma_{\rm lin}^{2}(R_{\rm ini}) & \xi_{\rm lin}(R_{\rm ini},r) \\
\xi_{\rm lin}(R_{\rm ini},r)  & \sigma_{\rm lin}^{2} (R_{\rm ini})
\end{array}
\right)\,,
\end{equation}
where the linear variance is given by~\eqref{eq:variance} and the correlation function by~\eqref{eq:corrlin}.
This covariance matrix can be diagonalised by transforming $(\tau,\tau')$ to a set of independent variables $(\nu,\zeta)$
\begin{align}
\label{eq:decorr}
\nu&=\frac{\tau}{\sigma}\,,\quad \zeta= \frac{ \sigma^2 \tau'-\xi\tau}{\sigma \sqrt{ \sigma^{4}-\xi}}\,,
\end{align}
which are built to be decorrelated $\left\langle \nu\zeta\right\rangle=0$ and have unit variance $\left\langle \nu^{2}\right\rangle=\left\langle \zeta^{2}\right\rangle=1$.
For Gaussian initial conditions, thanks to  diagonalization, $(\nu,\zeta)$ now follow a standard normal distribution, such that it is easy to check that the sphere bias reads
\begin{align}
b^{\rm G}_{\rm ini}(\tau)=\frac{\left\langle \tau'(\zeta,\nu)|\nu=\tau/\sigma\right\rangle}{\xi(r)}
=\frac{\tau}{\sigma_{\rm lin}^{2}(R_{\rm ini})}\,,
\label{eq:biasG}
\end{align}
which is proportional to the initial overdensity $\tau$ as expected from \cite{Kaiser84}. Furthermore, it is independent of $r$ such that the separation and density dependence in equation~\eqref{eq:jointPDFlargesep} can be factorised. 

\subsubsection{Non-Gaussian initial conditions: cumulant expansions}

For initially non-Gaussian fields, one has to redo the computation for the conditional mean that defines the initial bias function, equation~\eqref{eq:Kaiserbiasini}, which now might acquire a dependence of the separation that was not present for Gaussian initial conditions.
To obtain an expression for the two-point correlation for initially non-Gaussian fields,  the joint cumulant generating function was  expanded following  equation~\eqref{eq:cumgen}. The leading order correction to a bivariate Gaussian PDF will be the first term of a bivariate Edgeworth expansion \citep{Chambers67,McCullagh84,Kendall87} given in terms of third order cumulants, the skewness $\kappa_3=\langle\delta^3\rangle$ given in equation~\eqref{eq:skewnesslocal} and the mixed cumulant $\kappa_{12}=\langle\delta'\delta^2\rangle=\kappa_{21}$ given in equation~\eqref{eq:mixedskewness}\footnote{Note that this treatment resembles what is performed in \cite{Silk11}, where the scale-dependence of the bias is   computed based on a bivariate Edgeworth expansion for models with pure $g_{\rm NL}$.}.  In the decorrelated variables $\nu$ and $\zeta$ introduced in equation~\eqref{eq:decorr} this expansion is given in terms of products of Hermite polynomials \citep[see e.g.][]{BarndorffNielsen79Edgeworth}, which for the first non-Gaussian correction gives
\begin{align}
\label{eq:bivariateEdgeworth1st}
\mP(\nu,\zeta) &\approx \frac{\exp\left(-\frac{\nu^2}{2} - \frac{\zeta^2}{2}\right) }{2\pi} 
 \Bigg[ 1  \!+\! \frac{1}{6} \Big(\left\langle \nu^3 \right\rangle H_3(\nu) 
\!+\! \left\langle \zeta^3 \right\rangle H_3(\zeta) \Big)\notag \\ 
&\!+\! \hskip -0.2cm \frac{1}{2} \Big(\left\langle \nu \zeta^2 \right\rangle H_1(\nu) H_2(\zeta) 
\!+\! \left\langle \nu^2 \zeta \right\rangle H_2(\nu) H_1(\zeta) \Big) \Big]\,.
\end{align} 
with $H_1(x)=1, H_2(x)=x^2-1$ and $H_3=x^3-3x$. For an Edgeworth expansion up next-to-leading order that includes the joint kurtosis see Appendix~\ref{app:Edgeworthbias} and Appendix~\ref{app:NLOfnl} for the associated cumulants induced by $f_{\rm NL}^2$ and $g_{\rm NL}$. The conditional mean that determines the sphere bias is
\begin{equation}
\langle\tau' | \tau \rangle = 
\frac{ \displaystyle \! \int \! \! d\nu \! \int \! \! d\zeta \, \tau'(\nu,\zeta) \mP(\nu,\zeta) \delta_{\rm D}(\tau(\nu,\zeta) - \tau) }
{\displaystyle \! \int \! \! d\nu \! \int \! \! d\zeta \; \mP(\nu,\zeta) \delta_{\rm D}(\tau(\nu,\zeta) - \tau) }\,, \nonumber
\end{equation}
where inverting equation~\eqref{eq:decorr} gives
\begin{equation}
\tau= \nu \sigma ~, \quad 
\tau' = \frac{ \nu \xi +  \zeta \sqrt{\sigma^4 - \xi^2}}{\sigma}\,.
\end{equation}
After some algebra and using the properties of Hermite polynomials,
and expressing the moments of $\nu, \zeta$ back via the moments of $\tau, \tau'$, 
the sphere bias at leading order in $f_{\rm NL}$ becomes
\begin{align}
\label{eq:KaiserbiasnonGini}
b^{\rm NG}_{\rm ini}(\tau,r) &= \frac{\tau}{\sigma^{2}} \left[1+\frac{1}{2} \frac{\tau}{\sigma^{2}} \left(\frac{\kappa_{12}(r)}{\xi(r)} - \frac{\kappa_3}{\sigma^2}\right) \right]\\
\notag  &\qquad \qquad  -\frac{1}{2\sigma^2} \left(\frac{\kappa_{12}(r)}{\xi(r)} - \frac{\kappa_3}{\sigma^2}\right) \,,
\end{align}
where the cumulants $\kappa_3$, $\kappa_{12}$, $\sigma$ and $\xi$ from equations~\eqref{eq:skewnesslocal},~\eqref{eq:3ptcorr},~\eqref{eq:variance} and \eqref{eq:corrlin} are computed with smoothing radius $R_{\rm ini}$ and the term in the second row ensures a zero mean bias $\langle b(\tau)\rangle=0$. For an analogous result for a pure $g_{\rm NL}$ model see equation~\eqref{eq:KaiserbiasnonGinignl} in Appendix~\ref{app:Edgeworthbias} and for the generalisation of cumulants to arbitrary primordial bispectra see Appendix~\ref{app:beyondlocal}. In contrast to the result for Gaussian initial conditions, the result is now separation-dependent and the average two-point correlation function explicitly appears. Let us observe that the non-Gaussian sphere bias is a sum of the Gaussian sphere bias and a separation-dependent correction term
\begin{align}
b^{\rm NG}_{\rm ini}(\tau,r)=b^{\rm G}_{\rm ini}(\tau)+\Delta b^{\rm NG}_{\rm ini}(\tau,r)\,.
\end{align}
This expression shows that in analogy to the strong scale-dependent bias for overdense regions (halos), one expects an analogous scale-dependent bias for underdense regions (voids). This could be a promising idea for constraining pNG; while it has been mentioned in \cite{SekiguchiYokoyama12withdrawn}, no exhaustive study of this effect has been performed yet, but is underway\footnote{private communication with Hamaus, N. \& Chan, K.}. 

\subsection{Two-point sphere bias for an evolved non-Gaussian field}  
The two-point sphere bias for the time-evolved density field is defined in complete analogy to equation~\eqref{eq:Kaiserbiasini} as 
\begin{equation}
\label{eq:spherebiasdef}
b_R(\rho,r)=\frac{\langle \rho'|\rho;r\rangle-1}{\langle \tilde\rho\rho'; r\rangle-1} \,,
\end{equation}
where the numerator is the conditional mean sphere density given a sphere density $\rho$ at separation $r$ and the denominator is the average sphere correlation function. As was done for the one-point PDF,  a replacement rule for the initial density and radius is used following the spherical collapse~\eqref{eq:SCreplacement}, and the variance is rescaled according to equation~\eqref{eq:variancerescaling}. For Gaussian initial conditions and at large separations, this recovers the separation-independent expression given in \cite{Uhlemann17Kaiser}
\begin{align}
b^{\rm G}_R(\rho)= \frac{\sigma_{\rm lin}^2(R)\tau_{\rm SC}(\rho)}{\sigma^2(R)\sigma_{\rm lin}^2(R\rho^{1/3})}\,,
\end{align}
where the redshift dependence is encoded in the nonlinear variance $\sigma^2(z,R)$ and hence bias grows like $D^{-2}(z)$.
The pNG-induced extra terms of equation~\eqref{eq:KaiserbiasnonGini} can be rewritten in terms of reduced cumulants giving the combination $\tilde C_{12}-\tilde S_3$ where $\tilde C_{12}=\kappa_{12}/(\xi\sigma^2)$. Those reduced cumulants are robust against nonlinearities such that no further rescaling is needed. Due to the power of (linear) densities appearing in the reduced cumulants, the redshift-dependence of this term is $\propto D^{-1}(z)$.
For Gaussian initial conditions, the sphere bias is scale-independent, meaning that there is no residual dependence on pair separation. For non-Gaussian initial conditions the situation is different and one needs to relate separation dependent quantities at early and late times, more precisely the initial (Lagrangian) and final (Eulerian) separation of the two spheres (of identical density). \cite{Valageas10} suggest to treat each sphere as a test particle that falls into the potential well caused by another sphere and hence to use the mass conservation mapping for both positions to obtain
\begin{align}
\label{eq:separationreplacement}
r_{\rm ini}\simeq r\left(1+\frac{2}{3}\tau_{\rm profile}(\rho,R_{\rm ini},r)\right) \,,
\end{align}
where $\tau_{\rm profile}$ is the linear density contrast within radius $r$ around the sphere of radius $R_{\rm ini}$ with density $\rho$. It probes the radial density profile of the spherical saddle point which has been obtained in \cite{Bernardeau94rareevents,Valageas09}
\begin{align}
\label{eq:densityprofile}
\tau_{\rm profile}(\rho,R_{\rm ini},r)=\tau_{\rm SC}(\rho)\frac{\sigma_{\rm lin}(R_{\rm ini},r)^2}{\sigma_{\rm lin}(R_{\rm ini})^2} \,.
\end{align}
For large separations $r\gtrsim 60$Mpc$/h$ and the range of radii $R\simeq 15$ Mpc$/h$ and densities of interest here $\rho\in[0.5,3]$, the density contrast according to this profile is at the 5-10\% level. While there are in principle modifications of the density profile in the presence of pNG \citep[as calculated in][]{Valageas10}, their effect is negligible to leading order. Altogether the non-Gaussian bias is now given as
\begin{align}
\label{eq:KaiserbiasnonG}
b^{\rm NG}_R(\rho,r) &= b^{\rm G}_R(\rho)\left[1+
\frac{\tau_{\rm SC}(\rho)}{2} (\tilde C_{12} - \tilde S_3)(R_{\rm ini},r_{\rm ini}) \right]\,,
\end{align}
where the result is normalised to ensure $\langle b^{\rm NG}_R(\rho,r)\rangle=0$ and $\langle\rho b^{\rm NG}_R(\rho,r)\rangle=1$ as follows
\begin{equation}
\label{eq:KaiserbiasnonGnorm}
\hat b^{\rm NG}_R(\rho,r)=\frac{b^{\rm NG}_R(\rho,r)-\langle b^{\rm NG}_R(\rho,r)\rangle}{\langle (\rho-1)b^{\rm NG}_R(\rho,r)\rangle} \,.
\end{equation}

\subsubsection{Scale-dependent halo bias in Fourier space}
In Fourier space, scale-dependent halo bias has been derived using several approaches, as  summarised in \cite{Grossi09}; this  includes peak theory \citep{Dalal08}, the bispectrum \cite{MatarreseVerde08}, peak-background split \citep{Slosar08}, ellipsoidal collapse \citep{AfshordiTolley08}, perturbation theory \citep{Taruya08,McDonald08} and excursion sets \citep{Adshead12}. All of them lead to consistent results for the scale-dependent correction to the linear bias factor induced by leading order local non-Gaussianity
\begin{align}
\label{eq:biasnonG}
b_{\rm h, NG}(k) &= b_{\rm h,G} + \Delta b_{\rm h, NG}(k)\,,\\
\Delta b_{\rm h,NG} (k) &=2(b_{h,\rm G}-1)f_{\rm NL} \delta_c\frac{3\Omega_m}{2D(z)r_H^2k^2}\,,
\end{align}
where $r_H$ is the Hubble radius, $\delta_c$ the critical threshold for collapse and $D(z)$ the growth rate. A benchmark of theoretical predictions against numerical simulations has been performed in \cite{Desjacques09,Pillepich10,Giannantonio10,Smith11}. The presence of an extra term, corresponding to the term proportional to the initial skewness $\epsilon_3$ in equation~\eqref{eq:KaiserbiasnonG}, has been noted in \cite{Slosar08,AfshordiTolley08}. While it was argued in \cite{Slosar08} that  this scale-independent contribution does not cause a problem when fitting the bias to large-scale structure data, \cite{Desjacques09} found that it  leads to substantial improvement when comparing theoretical prediction to simulations. 

In this context, the generalisation to other types of pNG can be done by promoting the constant $f_{\rm NL}$ to a scale-dependent function $f_{\rm NL}(k)\propto (kR_{L,h})^\Delta$ where $\Delta=0$ gives back local pNG, $\Delta=2$ corresponds to equilateral pNG and $\Delta\in [0,2]$ includes quasi-single field inflation and some of its modifications. For a discussion of scale-dependent bias for bispectrum templates beyond local pNG see e.g. \cite{Scoccimarro12nonlocalPNG,Gleyzes17}.

\subsubsection{Scale-dependent halo bias in real space}
In real space, which is more natural for the treatment of densities-in-spheres, the scale-dependence of halo bias has been derived in \cite{MatarreseVerde08} based on an expansion of the two-point correlation for peaks as regions above a certain (high) threshold using a path-integral approach \cite{GrinsteinWise86,Matarrese86} and in \cite{Valageas10} with a steepest decent approach for leading-order local non-Gaussianity $f_{\rm NL}$.

According to \cite{MatarreseVerde08}, the equal density two-point correlation for initially non-Gaussian fields in the limit of high peaks and large separations is given by the result for Gaussian initial conditions, $\xi_{\rm h,G}=\xi_{\rm h}(r)b^{h,\rm G}_{\rm ini}(\tau)^2$, and a correction term, also called scale-dependent bias, is given by
\begin{align}
{\Delta\xi}^{\rm NG}_{\rm h}(r_{\rm ini},\tau) 
=\kappa_{12}(r_{\rm ini}) b^{\rm G}_{\rm ini}(\tau)^3 \,,
\end{align}
with $r=|\vx_2-\vx_1|$ and  with the leading order mixed cumulant $\kappa_{12}(r)$ given by equation~\eqref{eq:3ptcorr}. Hence, this yields to a bias function with only the first two terms in equation~\eqref{eq:KaiserbiasnonGini}.
However, following  the arguments in \cite{Valageas10}, one has to include another (separation-independent) term proportional to $\epsilon_3\sigma$ that stems from the skewness of the initial PDF and leads to the first line of \eqref{eq:KaiserbiasnonGini}, but is missing the constant terms.

As explained in \cite{Desjacques09}, a contribution from the non-Gaussian modification to the (linear) power spectrum is also expected; it is however much smaller than the terms presented before and will therefore be neglected, as was done in \cite{MatarreseVerde08} and \cite{Valageas10}.

\subsection{Phenomenological effect of primordial non-Gaussianities}
\label{subsec:phenopNG}

\begin{figure}
\hspace{-1.2cm}\includegraphics[width=1.2\columnwidth]{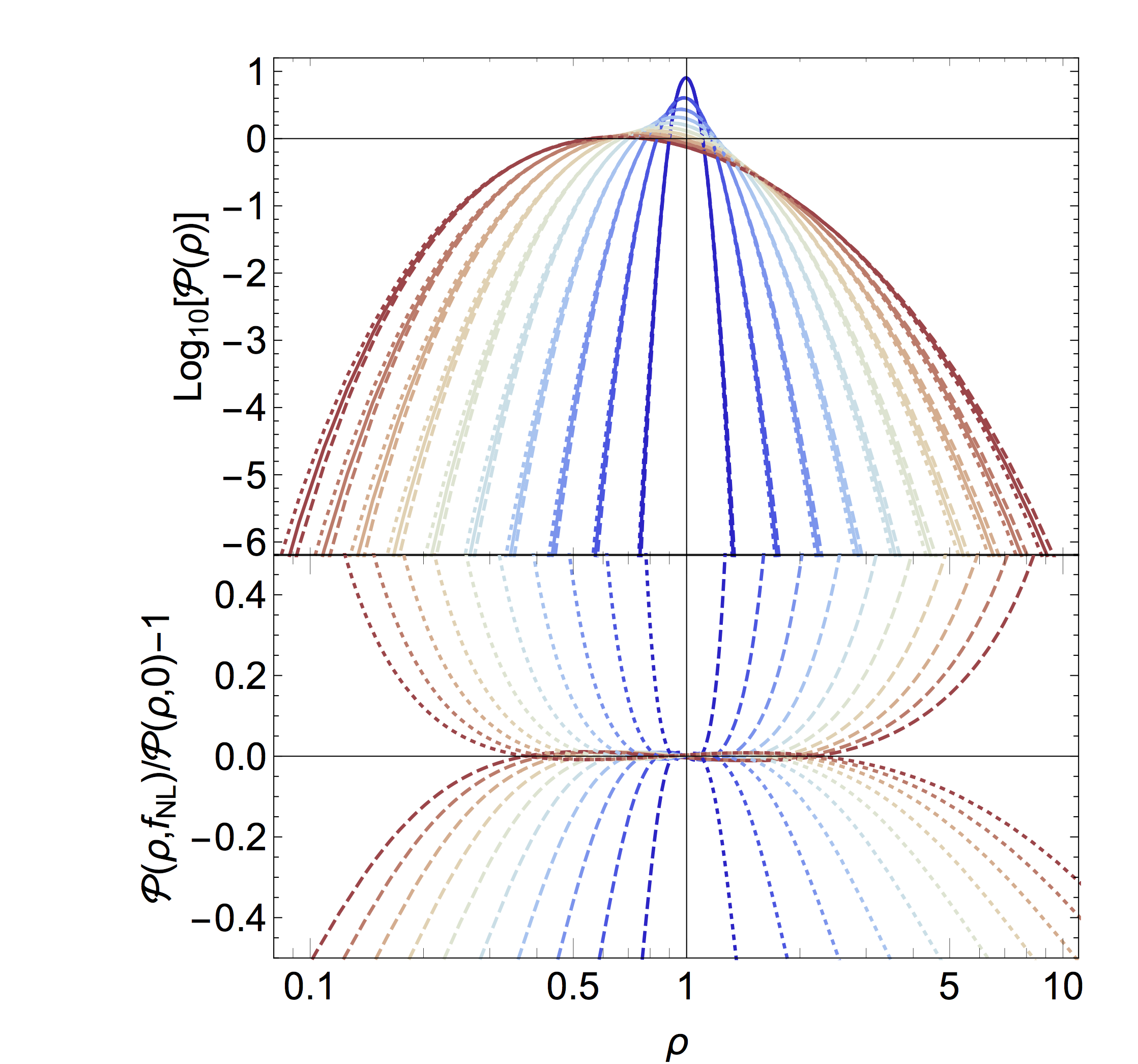}
\caption{(upper panel) 
The PDF of the log density $\mu=\log\rho$ in a sphere of radius $R=15$ Mpc$/h$ comparing a Gaussian model (solid lines) with one with local primordial non-Gaussianity according to equation~\eqref{eq:lpnG} for the gravitational potential $\Phi_p$ with $f_{\rm NL}^\Phi =\pm 100$ (dashed and dotted lines) for variances from $\sigma_\mu=0.05$ (blue) to $0.5$ (red) in steps of $0.05$ (lower panel) The corresponding ratio of the PDFs with and without primordial non-Gaussianity as a function of the density.
}  
\label{fig:PDFfnl100}
\end{figure}

\begin{figure}
\includegraphics[width=1.0\columnwidth]{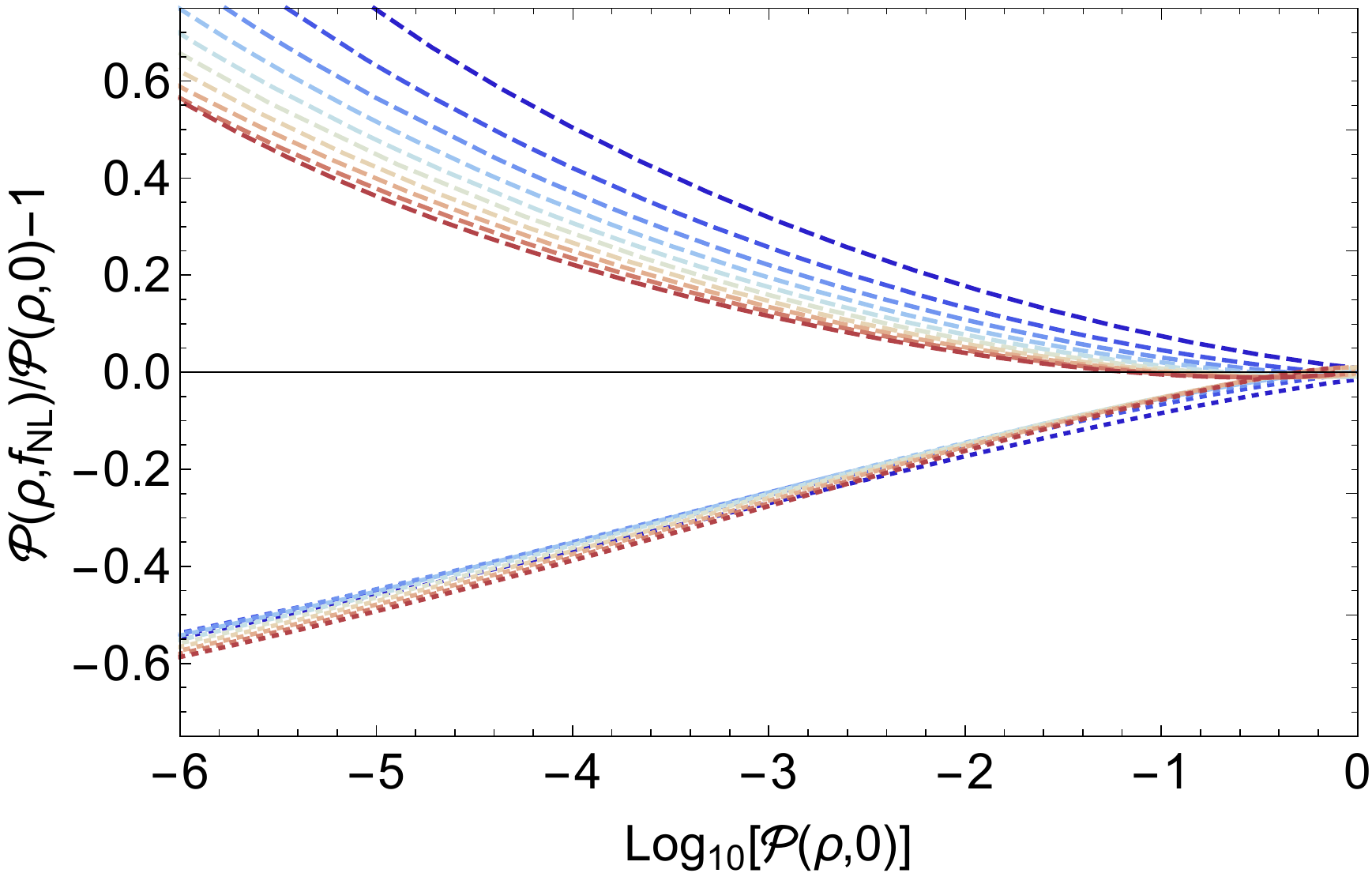}
\caption{The ratio of the one-point PDFs with $f_{\rm NL}=0$ and $f_{\rm NL}=+100$  for variances from $\sigma_\mu=0.05$ (blue) to $0.5$ (red) in steps of $0.05$ as a function of the rarity of the event for overdensities (dashed upper lines) and underdensities (dotted lower lines).
}  
\label{fig:PDFfnl100rarity}
\end{figure}

Figure~\ref{fig:PDFfnl100}  shows a comparison of the theoretical predictions for a fixed value of $f_{\rm NL}=100$ when varying the underlying nonlinear variance in a way that corresponds to time evolution $\sigma(z)=\sigma_{15}\cdot D(z)$ with $\sigma_{15}\simeq 0.5$ and  $D(z)$ the linear growth function. Note that although the nonlinear variance does not agree with the linearly evolved variance at late times, the late-time nonlinear  variance on the scales of interest grows approximately with the linear growth function. One can see that  the overall effect of primordial non-Gaussianity in the PDF is very weak except for the tails of the distribution. Note also that the apparent asymmetry between over and underdensities hints at a deviation from log-normality for high variances. 
Figure~\ref{fig:PDFfnl100rarity} shows the relative effect of pNG on the PDF at early (low variances, blue curves) and late times (high variances, red curves) when fixing the probability of the to-be-found density value and looking at overdensities (dashed upper lines) and underdensities (dotted lower lines). Note the three properties
\begin{enumerate}
\item For low variances corresponding to early times, the effect of pNG is significantly stronger for overdensities.
\item For high variances corresponding to late times, the effect of pNG is slightly stronger for underdensties.
\item The time (or variance) dependence is very mild for underdensities but substantial for overdensities. This is in line with the usual claim that underdensities are more pristine objects and better preserve the initial conditions.
\end{enumerate}
While those observations are interesting from a theory point of view, they should not be naively translated to practical applications for which several complications (in particular possible inaccuracies of the PDF in the tails, the limited available volume and a discrete sampling of the density field in observations) need to be considered.

\begin{figure}
\hspace{-0.5cm}
\includegraphics[width=1.1\columnwidth]{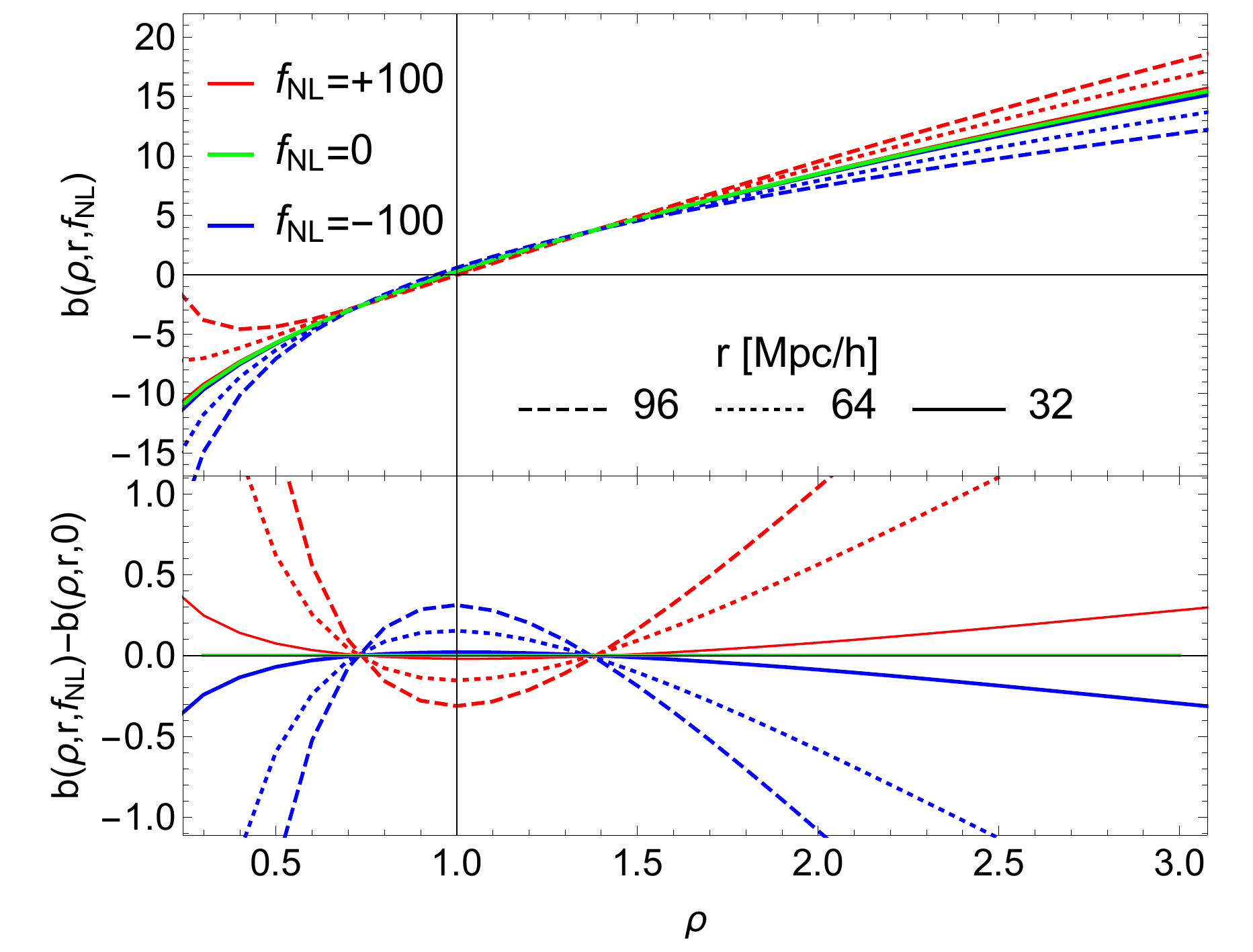}
\caption{Density-dependence of the scale-dependent two-point sphere bias $b_R(\rho,r)$ at redshift $z=1$ with radius $R=15$ Mpc$/h$ (where $\sigma_\mu\simeq 0.31$) and separations $r=32,64,96$ Mpc$/h$ (solid, dotted, dashed lines) induced by primordial non-Gaussianity with $f_{\rm NL}=\pm 100$ (blue and red lines, respectively). }
\label{fig:scaledepbias}
\end{figure}

 Figures~\ref{fig:scaledepbias}  displays  a schematic picture of the scale-dependence of the sphere bias  $b_R(\rho,r)$ that is induced by primordial non-Gaussianity and describes density-dependent clustering. The scale-dependent sphere bias correction given in equation~\eqref{eq:KaiserbiasnonG} scales like $\Delta b_R(\rho,r,f_{\rm NL}) \propto f_{\rm NL} b_{R,\rm G}(\rho) r^2$ such that it increases quadratically with separation $r$ (see Figure~\ref{fig:scaledepxiratiomodels}), the bias of the density $\rho$ in the sphere (hence the rarity of the event) and the amplitude of primordial skewness $f_{\rm NL}$. As dicussed before, the redshift dependence of scale-dependent spheres bias correction is roughly $\Delta b_R\propto 1/D(z)$.

\section{Validation  on simulations}
\label{sec:validation}

Let us now compare the prediction of equation~\eqref{eq:saddlePDFlog} to density in spheres measurements in 
dark matter simulation encoding primordial non-Gaussianities.

\subsection{Simulation}
The simulations contain $2048^3$ particles in a box of length $4096$ Mpc$/h$ and a given realization  has been run for models with $f_{\rm NL}=+100,0,-100$. The relevant cosmological parameters of the simulations (based on WMAP5) are summarized in Table~\ref{tab:cosmoparam} and snapshots are created at redshifts $z=3,2,1,0.35$. The non-Gaussian initial condition generator was developed in \cite{Nishimichi12} based on a parallel code by \cite{ValageasNishimichi11}, which computes the particle displacements using the second-order Lagrangian perturbation theory \citep[e.g.][]{Scoccimarro98,Crocce06}. The gravitational evolution is followed using the Tree-PM code Gadget2 \citep{gadget2,gadget}.
\begin{table}
\begin{tabular}{c|c|c|c|c|c}
\T
$\Omega_m$ & $\Omega_\Lambda$ & $h$ & $A_s(k_0=0.002/h$Mpc$^{-1})$ & $n_s$ & $\sigma_8$\\\hline
\T
0.279 & 0.721 & 0.701 & 2.486$\cdot 10^{-9}$ & 0.96 & 0.8157
\end{tabular}
\caption{Cosmological parameters of the simulation}
\label{tab:cosmoparam}
\end{table}

Table~\ref{tab:variance}  provides the values for the variance, the driving parameter of the theory, measured in the simulation for both the log-density $\mu=\log\rho$ and the density $\rho$. The differences in the variances measured at $R=15$ Mpc$/h$ are at the sub-percent level which is in qualitative agreement with the finding in \cite{Mao14}.
\begin{table}
\centering
\renewcommand{\arraystretch}{1.4} 
\setlength{\tabcolsep}{3pt}
\begin{tabular}{c|ccc|ccc|ccc}
$z$ & \multicolumn{3}{c|}{0.35} & \multicolumn{3}{c|}{1.00} & \multicolumn{3}{c}{2.00} \\\hline
$f_{\rm NL}$ & -100 & 0 & +100 & -100 & 0 & +100 & -100 & 0 & +100 \\ \hline\hline
$\hat \sigma_\mu^{\rm PDF}$ & 0.419 & 0.418 & 0.417 & 0.315 & 0.314 & 0.313 & \multicolumn{3}{c}{0.218}\\\hline
$\hat \sigma_\mu$ & 0.415 & 0.414 & 0.413 & 0.312 & 0.312 & 0.311 & \multicolumn{3}{c}{0.218}\\
$\hat S_3^\mu$ & 0.231 & 0.294 & 0.358 & 0.198 & 0.286 & 0.374 & 0.148 & 0.277 & 0.406 \\
-$\hat S_4^\mu$ & 0.421 & 0.367 & 0.293 & 0.529 & 0.461 & 0.354 & 0.631 & 0.547 & 0.376\\\hline
$\hat \sigma_\rho$ &  0.441 & 0.442 & 0.444 & 0.323 & 0.323 & 0.324 & \multicolumn{3}{c}{0.222}\\
$\hat S_3^\rho$ & 3.31 & 3.40 & 3.5 & 3.24 & 3.34 & 3.45 & 3.16 & 3.30 & 3.45 \\ 
$\hat S_4^\rho$ & 19.6 & 20.9 & 22.3 & 18.4 & 19.8 & 21.4 & 17.3 & 19.2 & 21.2\\
\end{tabular}
\caption{Parameters of the simulation that characterize primordial skewness in terms of $f_{\rm NL}$ and the nonlinear variance
 of the log-density $\mu=\log\rho$ for $R=15$ Mpc$/h$ at redshifts $z=0.35,1,2$ as measured from the simulation and as fitted using the PDF template from equation~\eqref{eq:saddlePDFlog}. Note that differences in the nonlinear variance between different non-Gaussian models are sub-percent.
}
\label{tab:variance}
\end{table} 

The PDF was measured with a top-hat filter of radius $R=15$Mpc$/h$ using a FFT-based method with a clouds-in-cells mass assignment. The convergence against the mass assignment schemes (either nearest grid point  (NGP) 
or clouds-in-cells (CIC) with different number of grid points) has been tested yielding stable results for intermediate and large densities $\rho$ where the difference between the two schemes is less than 1\% and potential problems in the highly underdense regions with $\rho < 0.4$, where there is a difference between the results obtained with NGP and CIC. The results shown are based on a CIC mass assignment with $1280^3$ grid points, with both window and aliasing corrections implemented before multiplying the top-hat function in k-space.

To measure the sphere bias, encoding the excess correlation between densities in spheres according to equation~\eqref{eq:spherebiasdef}, a given separation $r$ is chosen, giving a grid of non-overlapping spheres. In that grid, the densities of the neighbouring spheres with right separation $r$ are then collected in bins of width $\Delta\rho=0.1$; precise formulas are given by equations (19) and (20) in \cite{Uhlemann17Kaiser}.

\subsection{Comparison of analytical predictions against simulations}

\subsubsection{One-point PDF}
\begin{figure}
\includegraphics[width=1.0\columnwidth]{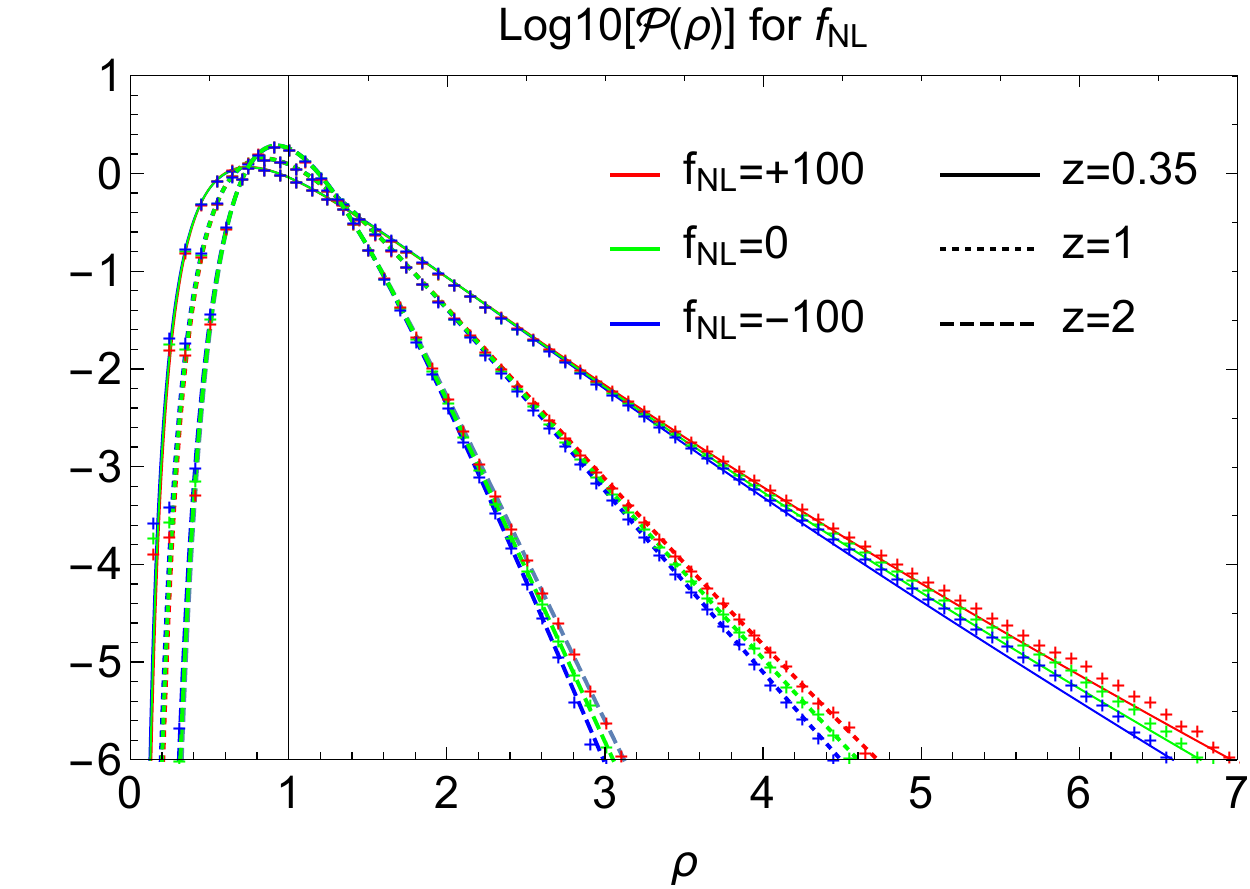}\\
\includegraphics[width=1.0\columnwidth]{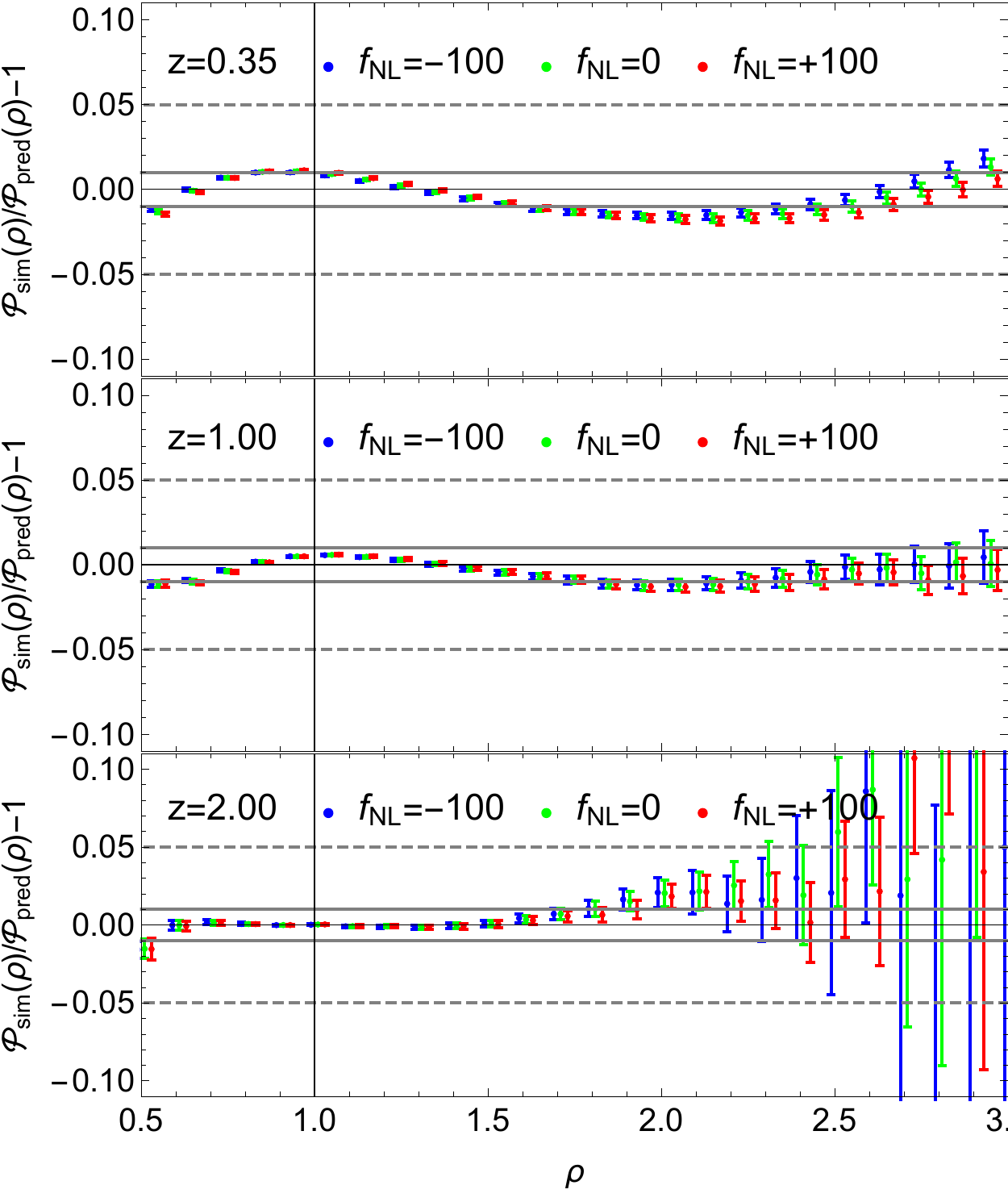}
\caption{(Upper panel) Comparison between measurement in the simulation (data points) for $f_{\rm NL}=-100,0,+100$ (blue, green, red) and the saddle point prediction (lines) at redshifts $z=0.35,1,2$ (solid, dotted, dashed) for the PDF with variances $\sigma_\mu^{\rm PDF}$ as given in Table~\ref{tab:variance}. (Lower panel) Corresponding residuals between the prediction and the measurements. For better visibility the $\rho$-values for $f_{\rm NL}=\pm 100$ have been displaced from the ones of $f_{\rm NL}=0$ by $\pm 0.02$. 
}  
\label{fig:compPDF_pred_sim}
\end{figure}

\begin{figure}
\includegraphics[width=1.0\columnwidth]{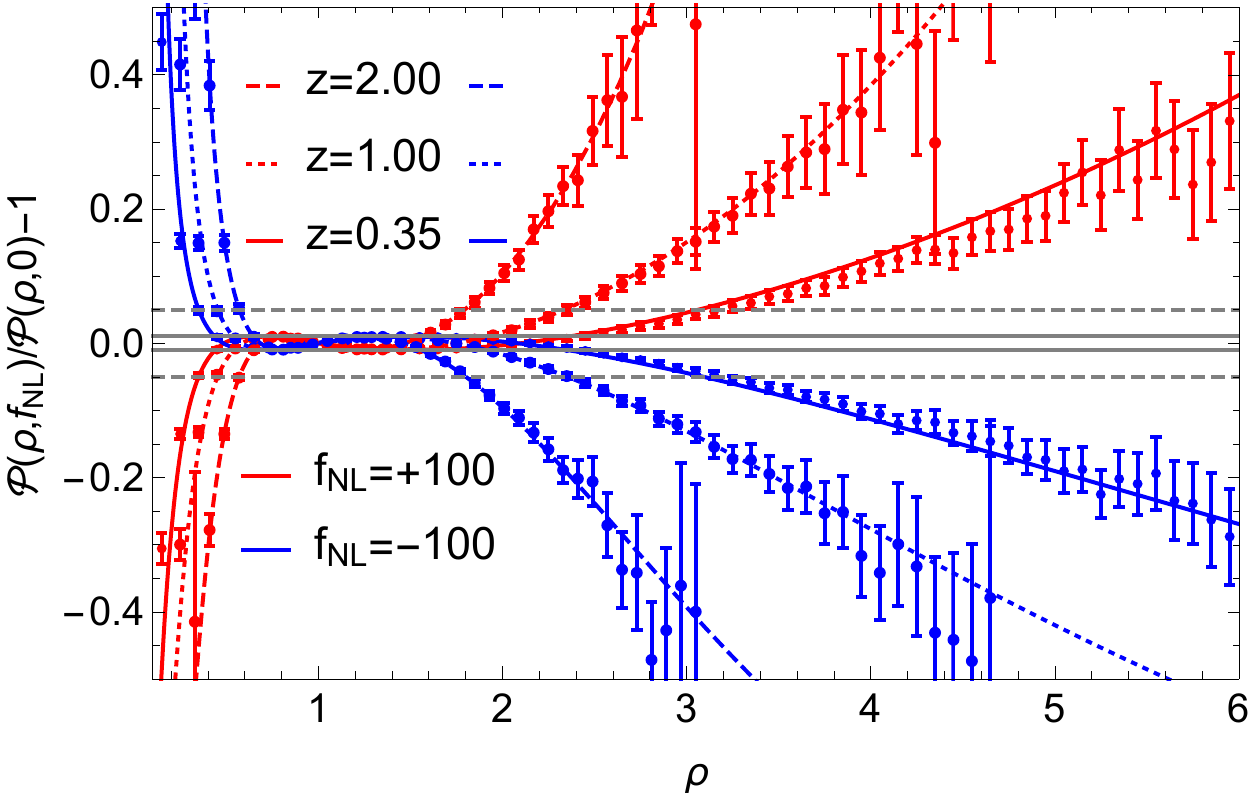}
\caption{Residuals between the PDFs with primordial non-Gaussianity $f_{\rm NL}=\pm 100$ (red and blue) and the fiducial PDF with $f_{\rm NL}=0$ as measured in the simulation (data points with error bars) and predicted from the saddle point approximation (lines) at redshifts $z=0.35,1,2$ (solid, dotted, dashed). The grey horizontal lines indicate pNG-induced deviations from the fiducial PDF at the 1\% (solid) and 5\% (dashed) level.}  
\label{fig:compPDF_nonG_G}
\end{figure}

Figure~\ref{fig:compPDF_pred_sim} shows a comparison of the theoretical predictions for the one-point PDF $\mP_R(\rho)$ against the simulations when plugging in the correct values of $f_{\rm NL}$ and the measured nonlinear log-variance $\sigma_\mu$ that is given in Table~\ref{tab:variance}. It shows that the prediction is in excellent agreement with the measurements, in particular it is accurate at the 1-percent level between densities of 0.5 and 2.5 and therefore can access a mildly nonlinear regime that is inaccessible to perturbative methods. 
 Figure~\ref{fig:compPDF_nonG_G}  shows the ratio of the Gaussian and non-Gaussian PDFs as predicted and measured in the simulation; an exquisite agreement is found  that shows that the inaccuracy of our PDF template acts the same way on the Gaussian and non-Gaussian version and hence comparing their ratio is in even better agreement with the measurements than what one would naively expect from the accuracy of the PDF which degrades for densities beyond what is shown in Figure~\ref{fig:compPDF_pred_sim}.

\subsubsection{Two-point sphere bias}
\begin{figure}
\hskip -0.8cm \includegraphics[width=1.1\columnwidth]{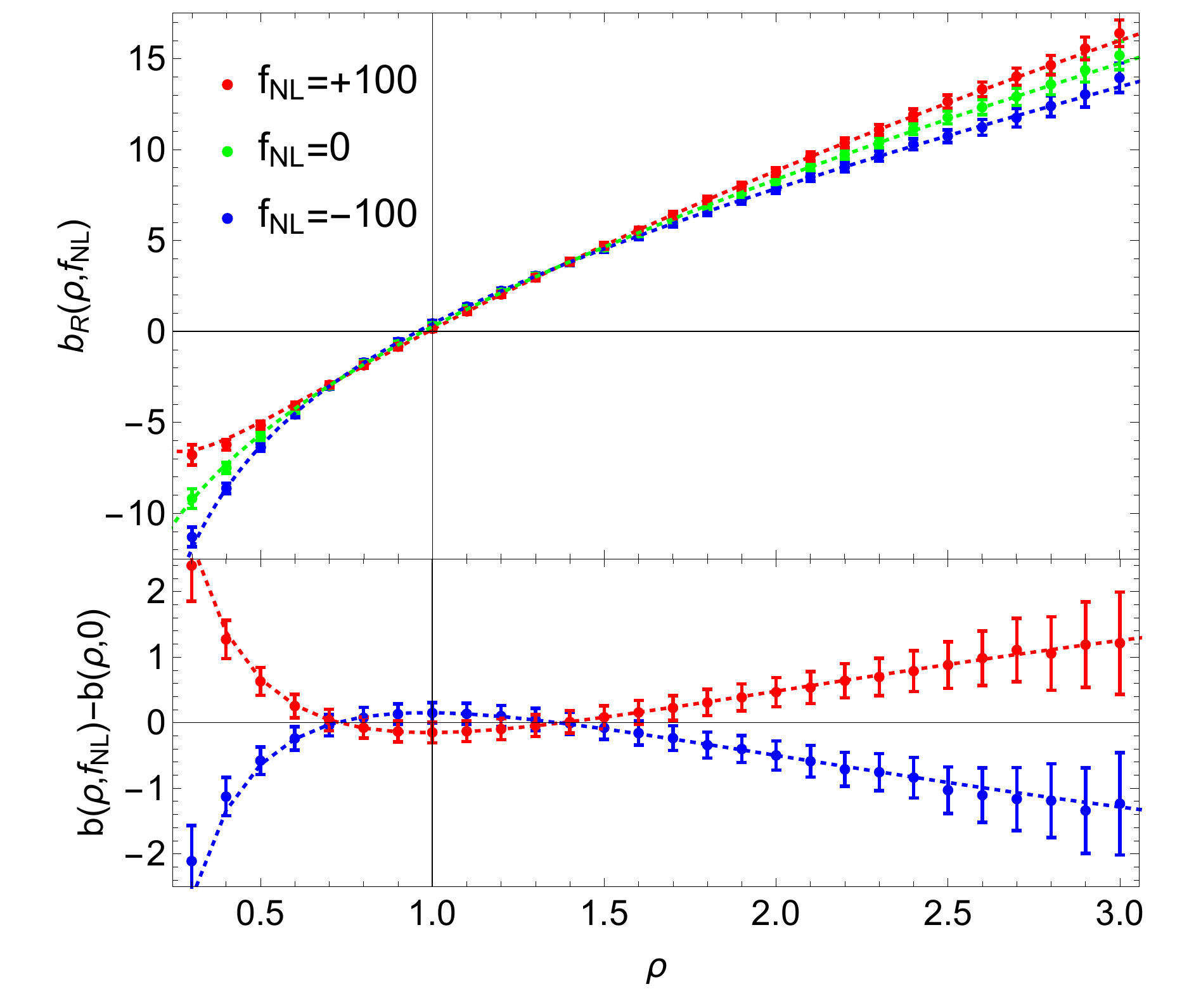}
\caption{Two-point sphere bias for local primordial non-Gaussianity $f_{\rm NL}=\pm 100$ (red and blue) and the fiducial PDF with $f_{\rm NL}=0$ (green) as measured in the simulation (data points with error bars) and predicted from the saddle point approximation (lines) at redshift $z=1$ for spheres of radii $R=15$ Mpc$/h$ at separation $r=64$ Mpc$/h$. 
}  
\label{fig:compspherebias}
\end{figure}

Figure~\ref{fig:compspherebias} shows a comparison of the theoretical prediction for the density- and separation-dependent two-point sphere bias $b_R(\rho,r)$. It appears clearly that the effect of primordial non-Gaussianity is strongest in the low and high density tails, in good agreement with the theoretical prediction from equations~\eqref{eq:KaiserbiasnonG}~and~\ref{eq:KaiserbiasnonGnorm}. The predictions shown are evaluated using the approximation $r_{\rm ini}\simeq r$ giving very similar results to the approximation from equation~\eqref{eq:separationreplacement} and only shows differences in the extreme density regions which are however within the error bars. The differences between the measurements with $f_{\rm NL}=\pm 100$ and $f_{\rm NL}=0$ indicating the impact of pNG on the clustering of certain density environments are in excellent agreement with the theoretical expectation. Note that, since the simulations have been run from the same realisation, error bars for the differences between are displayed as the individual error bars rather than their sum.

\section{Non-Gaussianity parameter estimators}
\label{sec:ML}

\subsection{Maximum likelihood (ML) with one-point statistics}

Having validated that the functional form of the one-point PDF (equation~\eqref{eq:saddlePDFlog}) against simulations, let us now use it to do a maximum likelihood estimate of the amount of primordial non-Gaussianity. The density PDF $\mP(\rho|\sigma, f_{\rm NL})$ at a given redshift depends on two parameters, the (gravitationally induced) nonlinear variance, $\sigma$, and the amplitude of primordial non-Gaussianity encoded in $f_{\rm NL}$. 

Let us  conduct the following fiducial experiment, following the assumptions  of \cite{Codis16DE} where the approximately linear growth of the nonlinear variance was used to constrain dark energy that changes the growth of fluctuations. Here, one can also take advantage of the fact that the nonlinear variance, when measured for spheres of radii $R>10$Mpc$/h$, evolves with the linear growth to a good approximation but keep the fiducial cosmology fixed except for allowing for primordial non-Gaussianity and changes in the overall amount of clustering. In order to mimic a Euclid-like survey, let us consider redshifts between $z_{\rm min}=0.6$ and $z_{\rm max}=2.0$\footnote{Limiting the redshift range to $z_{\rm max}=2$ also ensures that the effect of shot noise, which essentially cuts off the signal at high $z$ \citep[see e.g. Figure 3 in][]{Welling16} but is neglected here, is not too extreme.} binned so that the comoving distance of one bin is $d=40$Mpc$/h$. For every redshift bin   a number of spheres that corresponds to regularly drawn spheres of radius $R=15$Mpc$/h$ separated by $d=40$Mpc$/h$ are  considered. \footnote{At high redshifts,    the flat-sky approximation is not used;  instead  the volume of spherical shells of thickness $d$ is computed for all redshifts. To estimate the number of non-overlapping spheres that can be fit in the shells,  the average cut surface of the cosmic volume is divided by the cut surface of cubes with side length $d$. This estimate is close to the one obtained from assuming a hexagonal close packing with a density of $\eta=\pi/(3\sqrt{2})\simeq 0.74$ which is the densest possible packing of equal spheres.} For a 15,000 square degree survey, it yields 55 bins of redshift ($z_{i}$) with a number of spheres ranging from about $N_{1}\simeq7000$ (at $z_{1}=0.6$) to $N_{55}\simeq40,000$ (at $z_{55}=2.0$) for a total of about 1.2 million supposedly independent spheres.
In this experiment, note that samples are drawn directly from the PDF, hence assuming that the model for the density PDF is exact, meaning that the local amplitude $f_{\rm NL}$ well parametrises the amount of primordial non-Gaussianity in terms of skewness and non-linear variance $\sigma$ well describes the outcome of nonlinear gravitational evolution from close to Gaussian initial conditions.

Note that the present analysis is conservative with respect to the number of spheres one can get given a certain volume, but optimistic with respect to the modelling assuming no theoretical error. On the one hand, shrinking the radius $R$ of the spheres increases the available information and hence tightens constraints. On the other hand,  the accuracy of the theoretical prediction decreases with the radius such that the theoretical error becomes relevant. Hence a good balance of the two must be  struck. One possible improvement on the following analysis would be to decrease the radius of the spheres with increasing redshift (for example using linear growth) to keep the variance, which controls the theoretical accuracy, constant over the redshift slices. In practice, the chosen treatment to scale the number of drawn spheres with the available volume (and hence not probing the deep PDF tails, especially at low redshifts) ensures that the PDFs entering the ML estimation are within their regime of validity (as determined from the comparison with simulations).

\subsubsection{Estimate of primordial skewness}
In order to get constraints on the parameter $f_{\rm NL}$ of primordial non-Gaussianity, let us compute the log-likelihood of 10 randomly samples for the 1.2 million measured densities $\{\rho_{i,j}\}_{{1\leq i\leq N_{j}},{1\leq j\leq 55}}$ given models for which $f_{\rm NL}$ varies
\begin{equation}
{\cal L}(\{\rho_{i,j}\}|f_{\rm NL})=\sum_{j=1}^{55}\sum_{i=1}^{N_{j}}\log {\cal P}(\rho_{i,j}|z_j,f_{\rm NL})\,,
\end{equation}
where ${\cal P}(\rho|z_j,f_{\rm NL})$ is the theoretical density PDF at redshift $z_j$ for a primordial non-Gaussiniaty model parametrized by $f_{\rm NL}$. Optimizing the probability of observing densities $\{\rho_{i,j}\}_{{1\leq i\leq N_{j}},{1\leq j\leq 55}}$ at redshifts $\{z_{j}\}_{{1\leq j\leq 55}}$ with respect to  $f_{\rm NL}$, yields a maximum likelihood estimate for the primordial non-Gaussian parameter.  

The resulting mean maximum likelihood values (averaged over 10 samples) and the corresponding $\alpha=1,2,3$ sigma confidence intervals given in Table~\ref{tab:MLfnl} correspond to the models for which ${\cal L}(\{\rho_{i,j}\}|f_{\rm NL})=\max_{f_{\rm NL}} {\cal L}(\{\rho_{i,j}\}|f_{\rm NL}) +\log \left(1-\rm{Erf}(\alpha/\sqrt 2)\right)$.
Modulo our assumptions, this maximum likelihood method  allows to detect non-Gaussianity with $\sigma(f_{\rm NL})\simeq 10$.

\begin{table}
\centering
\renewcommand{\arraystretch}{1.4} 
\begin{tabular}{c|c|cccc}
 & ${\rm fid}$ & $\rm ML$ & 1$\sigma$ & 2$\sigma$ & 3$\sigma$
\\\hline
$f_{ \rm  NL}$ & 0.0 & -2.5 & $\pm 9.5$ & $\pm 16.5$ & $\pm 24.5 $\\
$\sigma_{15}$ & 0.51450 & 0.51445 & $\pm 0.0004$ & $\pm 0.0007$ & $\pm 0.001$\\
\end{tabular}
\caption{Collection of mean maximum likelihood (ML) results determined from 10 samples for $f_{\rm NL}$ parametrizing the amount of primordial skewness $f_{\rm NL}$ when keeping the nonlinear variance $\sigma_{15}$ fixed at its fiducial value and vice versa.}
\label{tab:MLfnl}
\end{table}

\subsubsection{Joint estimate of primordial skewness \&  variance}
\begin{figure}
\centering
\includegraphics[width=.9\columnwidth]{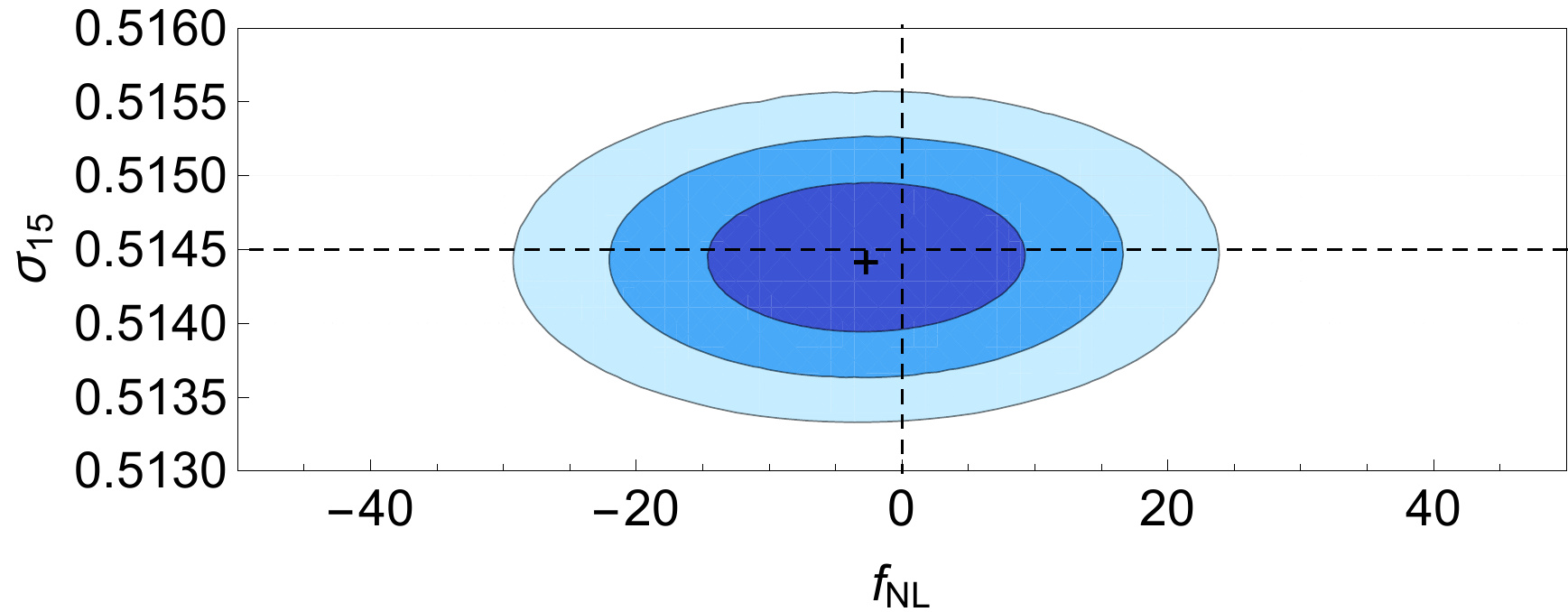}
\caption{Result of joint maximum likelihood for nonlinear variance $\sigma_{15}=\sigma(R=15,z=0)$ and primordial non-Gaussianity $f_{\rm NL}$ with the mean of the maximum (point) and the corresponding 1,2 and 3-$\sigma$ contours (dark blue to cyan) averaged over 10 samples drawn from the PDF at different redshift with fiducial values $(\sigma_{15},f_{\rm NL})=(0.5145,0)$ (dashed cross).
}  
\label{fig:jointML}
\end{figure}

In order to get  joint constraints on primordial skewness parametrized through $f_{\rm NL}$ and the nonlinear log-variance at given radius at present time $\sigma_{15}=\sigma_\mu(R=15\text{Mpc}/h,z=0)$ (hence equivalent to $\sigma_8$), let us re-compute the log-likelihood of the 1.2 million measured densities $\{\rho_{i,j}\}_{{1\leq i\leq N_{j}},{1\leq j\leq 50}}$ given models for which both $f_{\rm NL}$ and $\sigma_{15}$ vary
\begin{equation}
{\cal L}(\{\rho_{i,j}\}|\sigma_{15},f_{\rm NL})=\sum_{j=1}^{50}\sum_{i=1}^{N_{j}}\log {\cal P}(\rho_{i,j}|z_j,\sigma_{15}, f_{\rm NL})\,.
\end{equation}
 Figure~\ref{fig:jointML}  depicts the result for the joint maximum likelihood estimate computed for 10 different samples. It shows that the joint estimation of $f_{\rm NL}$ and $\sigma_{15}$ shows no degeneracy, hence the signature of primordial non-Gaussianity is qualitatively different from a change in the normalization of the density fluctuation field, associated to $\sigma_8$ (or in our case $\sigma_{15}$). Having access to both (overdense and underdense) wings  of the PDF  is what  helps distinguishing these two effects. The marginalization over $\sigma_{15}$ decreases the accuracy that can be obtained for $f_{\rm NL}$ only marginally compared to the run where  $\sigma_{15}$ was kept fixed at its fiducial value. 

\subsection{Observational effects}
\textbf{Tracer bias.} 
It has been demonstrated recently in \cite{UhlemannBias17} that tracer bias can be incorporated in densities-in-spheres statistics by a local mapping between average densities in spheres. Based on mass-weighted halo densities in both real and redshift space,  a local one-to-one mapping has been shown to provide accurate predictions for one- and two-point densities-in-spheres statistics. The relationship between dark matter and mass-weighted halo densities has been observed to admit a quadratic model in the log-densities as efficient parametrisation. The main effect of tracer bias is to cause a strong degeneracy between the dark matter variance and bias parameters for the one-point PDF which can only be lifted using two-point statistics. Since however, the dark matter variance has been shown not to be degenerate with a local $f_{\rm NL}$, bias is not expected to diminish the constraints except for finite sampling effects. This is in line with the finding in \cite{SefusattiKomatsu07} that for a bispectrum analysis, a local $f_{\rm NL}$ is not degenerate with (local) galaxy bias. In \cite{Mao14}, the signal of pNG in the first three one-point cumulants of the smoothed nonlinear density field has been studied  on numerical simulations and mock galaxy catalogs resembling LRGs. For dark matter densities, it was shown that the pNG signal in the skewness and kurtosis of the dark matter densities is much larger than that induced by the nonlinear variance. Note that in the present formalism, given that the one-point PDF includes all cumulants, the skewness and kurtosis are  both taken into account when jointly constraining the dark matter variance and $f_{\rm NL}$ via the maximum likelihood approach, which does improve the accuracy of the parameter estimation compared to sample variance \citep[as shown in][]{Codis16DE}. However, for number-weighted galaxy densities from  mocks, the finding in \cite{Mao14} is in stark contrast, and the variance almost completely dominates the pNG signal. Recently, \cite{Gleyzes17} investigated the scale dependent halo bias for equilateral and quasi-single field inflation generated pNG, and looked at  how they can or cannot be distinguished from a general biasing model. For local (and some quasi-single field) shapes, they found that significant improvements over Planck are achievable with  scale-dependent bias  for a large volume large-scale structure surveys, while the sensitivity to equilateral non-Gaussianity is heavily suppressed by marginalizing over a general halo biasing model.

\textbf{Redshift space distortions.}
The effect of redshift space distortions has been assessed in \cite{Mao14} where it has been found that they affect the variance and skewness very similarly in the Gaussian and non-Gaussian case and therefore do not affect the detectability of non-Gaussianity from these measurements, see Fig. 4 in \cite{Mao14}. Recently, \cite{UhlemannBias17} showed that redshift space distortions can be incorporated naturally in a mean local bias model for densities-in-spheres statistics such as one-point PDF and the two-point sphere bias.

\textbf{General relativistic (GR) effects.}
There are also GR effects that generate non-Gaussianity. According to \cite{Bruni14}, the nonlinear relation between the spatial curvature and the metric perturbation  translates into a specific non-Gaussian contribution to the initial comoving matter density. Note also that there are subtleties in the definition of bias depending on rest frames and relativistic effects can enter through projection or past light cone effects \citep[see Section 9 of][for a review]{Desjacques16Review}. Relativistic effects are most important on the very large scales that are relevant for the scale-dependent bias in Fourier space \citep{Camera15}, while the intermediate separations  probed here should not be affected as much. However, in \cite{Yoo12} it has been shown that the GR effect is not degenerate with the pNG signature in galaxy bias, and the ability to detect primordial non-Gaussianity is little compromised.

\subsection{Outlook}
For local pNG, one should eventually constrain primordial non-Gaussianity using jointly the one-point PDF and two-point sphere bias statistics. This could in particular be interesting for distinguishing between $f_{\rm NL}$ and $g_{\rm NL}$ which lead to a very similar signature in scale-dependent halo bias \citep{DesjacquesSeljak10gnl,Giannantonio10,DesjacquesJeongSchmidt11a,DesjacquesJeongSchmidt11b,Smith12}, see the extensions discussed in Appendix~\ref{app:details}. Indeed, joint analyses of abundances and clustering have been performed for galaxy clusters in \cite{Sartoris10,Mana13} and shown to improve constraints based on individual probes. To render densities-in-spheres statistics applicable to counts-in-cells measurements in galaxy surveys, one should furthermore include tracer bias and redshift space distortions.
While this is beyond the scope of this paper, the present procedure   to  include  primordial non-Gaussianity in densities-in-spheres statistics can be combined with the biased tracer  formalism described in \cite{UhlemannBias17} which also demonstrates a joint parameter estimation using the one-point PDF and the two-point sphere bias function. 

{Recently, \cite{Chiang15} introduced the position-dependent correlation function (or equivalently power spectrum) as the correlation between two-point functions of galaxy pairs within different (large) subvolumes with a given (small) mean density contrast at their location. Based on this, it has been demonstrated that for local $f_{\rm NL}$ the position-dependent correlation function can yield comparable constraints to the full bispectrum, despite its inability to distinguish between linear and quadratic bias. This is encouraging  because the spirit of the position-dependent correlation function is to capture the density-dependence of two-point clustering and hence similar to sphere bias which relies on different (small) spheres with a potentially large average density contrast.}

For nonlocal pNG such as the equilateral and orthogonal types, the expected signal in sphere bias is much smaller than in the local case, see Figure~\ref{fig:scaledepxiratiomodels}. Hence, one has to rely on the one-point PDF for which achievable constraints on $f_{\rm NL}$ should be of the same order than for the local case because the main qualitative change is in the scale-dependence of the skewness, see Figure~\ref{fig:scaledepskewnessmodels}.

\section{Conclusion}
\label{sec:conclusion}

Relying on analytical predictions for the one-point PDF and the two-point bias of spherically-averaged cosmic densities, we disentangled dynamically generated non-Gaussianity from primordial ones.
The  primordial non-Gaussianity is encoded in the form of a small initial skewness (controlled by $f_{\rm NL}$) with a scale-dependence that in general depends on the bispectrum of the underlying model and is very small for local pNG. 

We successfully benchmarked the analytical predictions for the one-point PDF from local non-Gaussianity in a range from $f_{\rm NL}=- 100$ to $+100$ against numerical simulations finding excellent agreement, achieving 1\% accuracy within the central density region $\rho\in[0.5,2.5]$ and about 5\% for the adjacent high/low density tails. Similarly, we have tested the predicted impact of pNG on two-point statistics encoded in the scale-dependence of the sphere bias against the simulation finding very good agreement within the error bars of the simulation. Most notably, we have observed a scale-dependent sphere bias in extreme density environments that includes overdensities, -- a  well-known result for halo bias -- but also underdensities suggesting that void bias could be used to constrain pNG as well.

Using a simple joint maximum likelihood estimator for the amplitude of the nonlinear variance $\sigma_8$ and local primordial skewness $f_{\rm NL}$, we obtained an estimate for the constraining power of the one-point density-in-spheres PDFs for a Euclid-like survey finding $\sigma(f_{\rm NL})\simeq 10$. We then discussed the influence of tracer bias, redshift space distortions and relativistic effects and provided an outlook for constraining pNG jointly using the one-point PDF and two-point bias while including bias. Given the clear scale-dependent bias signal we observed for both high and low densities spheres, one can hope to improve upon pNG constraints obtained from halo bias alone in the near future and better disentangle between $f_{\rm NL}$ and $g_{\rm NL}$.
\\
\\
{\bf Acknowledgements:}  
 This work is partially supported by the grants ANR-12-BS05-0002 and  ANR-13-BS05-0005 of the French {\sl Agence Nationale de la Recherche}. CU and EP are supported by the Delta-ITP consortium, a program of the Netherlands organization for scientific research (NWO) that is funded by the Dutch Ministry of Education, Culture and Science (OCW). CU thanks IAP for hospitality when this project was initiated. TN acknowledges financial support from JSPS KAKENHI Grant Number 17K14273 and JST CREST Grant Number JPMJCR1414.
Numerical simulations presented in this paper were carried out on Cray XC30 at Center for Computational Astrophysics, National Astronomical Observatory of Japan.

CU thanks Guido D'Amico, Matteo Biagetti, Kwan Chuen Chan, Vincent Desjacques, Nico Hamaus, Eiichiro Komatsu, Mikhail Ivanov, Marcello Musso, Marilena LoVerde, Mark Neyrinck, Emiliano Sefusatti, Fabian Schmidt, Ravi Sheth, Sergey Sibiryakov, Yvette Welling, Drian van der Woude for discussions and Toyokazu Sekiguchi together with Shuichiro Yokoyama for information.
  
 \newpage
\bibliographystyle{mnras}
\bibliography{LSStructure}

\appendix

\section{Beyond leading-order and local}
\label{app:details}
This appendix generalises the formalism presented for leading order local pNG to general primordial bispectra  and next-to-leading order local pNG. As prime example for arbitrary primordial bispectra, it will consider equilateral and orthogonal pNG and show how the model ingredients vary compared to local pNG. For next-to-leading order local pNG,  the clear hierarchy of primordial rescaled cumulants that is relevant for the validity of the Edgeworth-expansions for the primordial statistics used in the main-text will be exhibited.
\subsection{Scale dependence of primordial cumulants}
\label{app:gen}

\begin{figure}
\includegraphics[width=1.0\columnwidth]{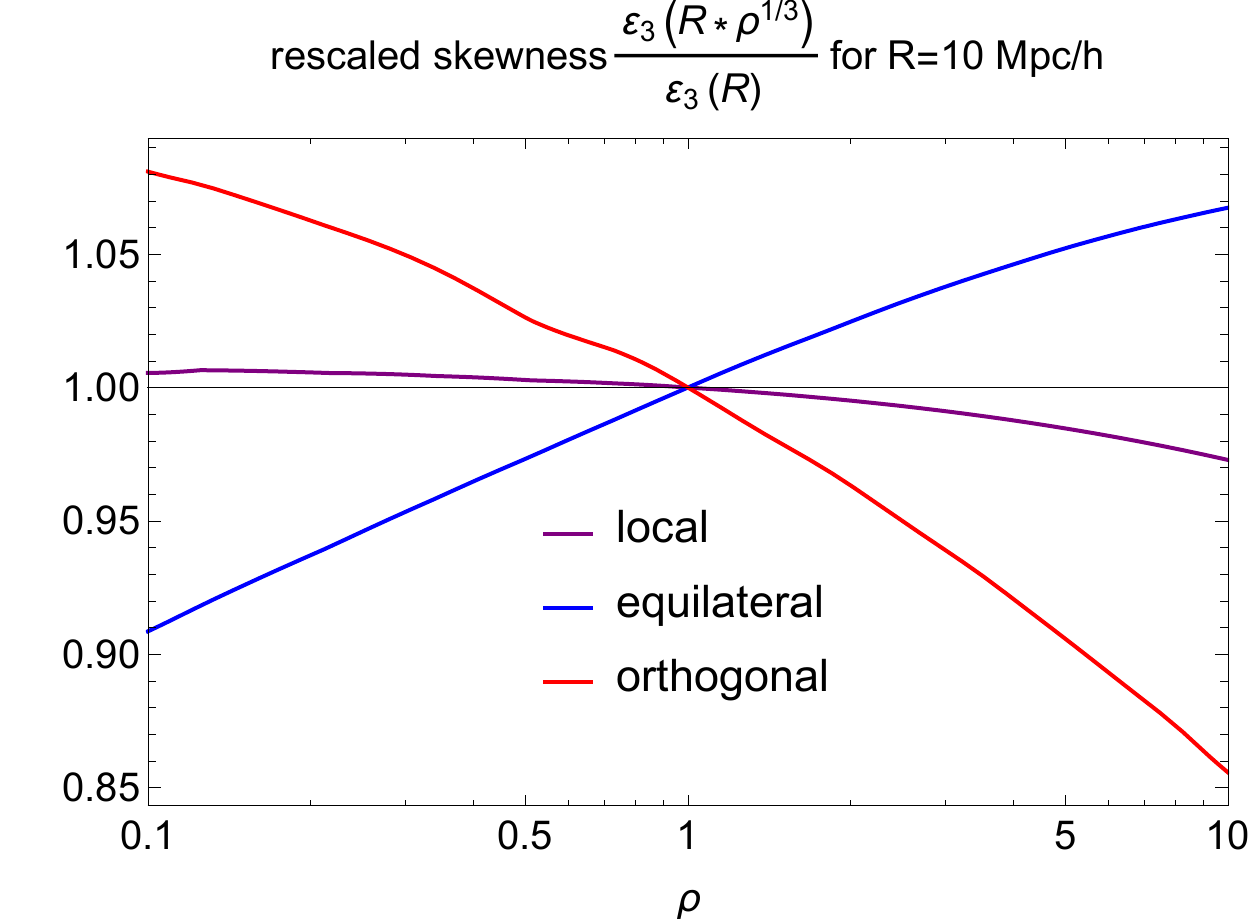}
\caption{Scale-dependence of the rescaled skewness for different primordial bispectra: local (purple), equilateral (blue) and orthogonal (red).  All models are close to having a scale-invariant skewness which is best achieved for a local model.}  
\label{fig:scaledepskewnessmodels}
\end{figure}

\begin{figure}
\includegraphics[width=1.0\columnwidth]{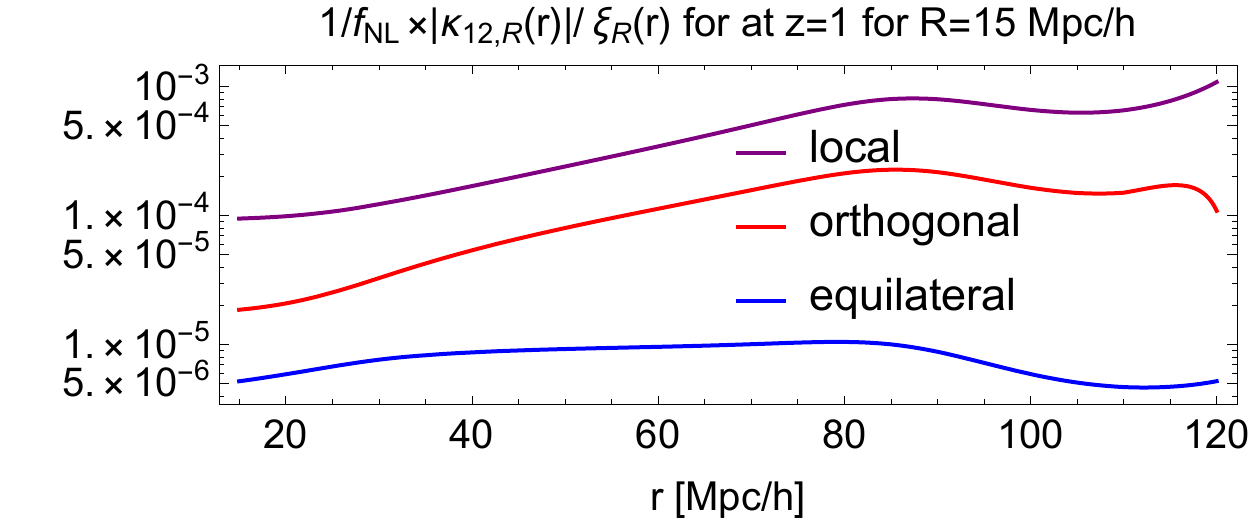}
\caption{Scale-dependence of the ratio of the leading order mixed cumulant $\kappa_{12,R}(r)$ and the linear correlation function $\xi_R(r)$ that enters the two-point sphere bias $b(\rho)$ evaluated at redshift $z=1$ with radius $R=15$ Mpc$/h$ as a function of separation $r$ [Mpc$/h$] normalised by the amplitude primordial non-Gaussianity given by $f_{\rm NL}$ for different primordial bispectra: local (purple), equilateral (blue) and orthogonal (red).}  
\label{fig:scaledepxiratiomodels}
\end{figure}

\subsubsection{Non-Gaussianity beyond local shape}
\label{app:beyondlocal}

Beyond the local non-Gaussianity, whose bispectrum is a simple product of two power spectra
\begin{align}
B^{\rm loc}_\Phi(\vk_1,\vk_2,\vk_3) &= 2f_{\rm NL}^{\rm loc} \left[P_\Phi(k_1)P_\Phi(k_2)+ \text{2 perm.}\right]\,,
\end{align}
there are many other shapes of the primordial bispectrum, for example of equilateral and orthogonal type
\begin{align}
\notag B^{\rm eq}_\Phi(\vk_1,\vk_2,\vk_3) &= 6f_{\rm NL}^{\rm eq} \Big[-[P_\Phi(k_1)P_\Phi(k_2)+ \text{2 perm.}]\\
& -2P_\Phi^{2/3}(k_1)P_\Phi^{2/3}(k_2)P_\Phi^{2/3}(k_3)\\
\notag &  +P_\Phi(k_1)P_\Phi^{2/3}(k_2)P_\Phi^{1/3}(k_3) + \text{5 perm.} \Bigg]\,,\\
\notag B^{\rm orth}_\Phi(\vk_1,\vk_2,\vk_3) &= 6f_{\rm NL}^{\rm orth} \Big[-3[P_\Phi(k_1)P_\Phi(k_2)+ \text{2 perm.}]\\
& -8P_\Phi^{2/3}(k_1)P_\Phi^{2/3}(k_2)P_\Phi^{2/3}(k_3)\\
\notag &  +3[P_\Phi(k_1)P_\Phi^{2/3}(k_2)P_\Phi^{1/3}(k_3) + \text{5 perm.}] \Big]\,.
\end{align}
Note that, in contrast to  the strictly local case, for the equilateral bispectrum there is no a priori reason for $f_{\rm NL}^{\rm eq}$ to be scale-independent \citep[see e.g.][]{LoVerde08}. While single-field inflationary models generically predict the local type,  single-field models generate predominantly other forms.
Local non-Gaussianity peaks for squeezed configurations $k_1\simeq k_2\ll k_3$, while equilateral peaks for $k_1\simeq k_2\simeq k_3$, note that a fair comparison of the amplitudes $f_{\rm NL}^{\rm loc}$ and $f_{\rm NL}^{\rm equi}$ requires careful normalization that takes the different shape-dependence into account,  as argued in \cite{FergussonShellard09}. 

Extending the leading order mixed cumulant given in equations~\eqref{eq:skewnesslocal}~and~\eqref{eq:mixedskewness} to a general form for the primordial bispectrum $B_\Phi$ yields
\begin{align}
\label{eq:skewnessbispectrum}
\kappa_{3,\rm NG}&= \! \int \! \frac{\dd^3k_1}{(2\pi)^3}\, \alpha_R(k_1) \! \int \! \frac{\dd^3k_2}{(2\pi)^3} \alpha_R(k_2)\\
\notag &\qquad\quad \alpha_R\left(|\vk_1+\vk_2|\right) B_\Phi(\vk_1,\vk_2,-(\vk_1+\vk_2))\,,
\end{align}
and
\begin{align}
\label{eq:mixedskewnessbispectrum}
\kappa_{12,\rm NG}(r)&= \! \int \!  \frac{\dd^3k_1}{(2\pi)^3} \alpha_R(k_1) \int\frac{\dd^3k}{(2\pi)^3} \alpha_R(k) \\
\notag &\qquad \quad \alpha_R(|\vk-\vk_1|) B_\Phi(\vk_1,-\vk_1-\vk,\vk) \exp[i \vk\cdot \vr]\,.
\end{align}
Since the present formalism does not make any assumption on the scale-dependence of the primordial cumulants, it can be applied to any known bispectrum shape and naturally incorporates models of fixed shape with running non-Gaussianity parameter as discussed in \cite{LoVerde08}. Note that for identical scale-independent prefactors $f_{\rm NL}^{\rm loc}=f_{\rm NL}^{\rm equi}$, the relative amplitude of the skewness differs $\kappa_3^{\rm loc} \simeq 3 \kappa_3^{\rm equi} \simeq -5 \kappa_3^{\rm orth}$ for radii around $10-15$ Mpc$/h$ while the density scale-dependence is shown in Figure~\ref{fig:scaledepskewnessmodels}. The separation scale-dependence of the leading order mixed cumulant is compared in Figure~\ref{fig:scaledepxiratiomodels}, which shows  that the orthogonal model has a sign opposite to the local and equilateral models. This figure shows that the signal of scale-dependent bias is largest for the local model, suppressed roughly by a factor 3 for the orthogonal model, but by one to two orders of magnitude for the equilateral model.

\subsubsection{Next-to-leading order local non-Gaussianity}
\label{app:NLOfnl}

One can also consider quadratic local primordial non-Gaussianity which can be written as
\begin{equation}
\label{eq:lpnGwithgnl}
\Phi_{\rm NG}= \Phi_G+ f^\Phi_{\rm NL} \left(\Phi_G^2-\langle\Phi_G^2\rangle\right) 
+ g^\Phi_{\rm NL} \left(\Phi_G^3-3\Phi_G\langle\Phi_G^2\rangle\right)\,,
\end{equation}  
with an extra constant $g^\Phi_{\rm NL} $ compared to \eqref{eq:lpnG}.
\begin{figure}
\includegraphics[width=1.0\columnwidth]{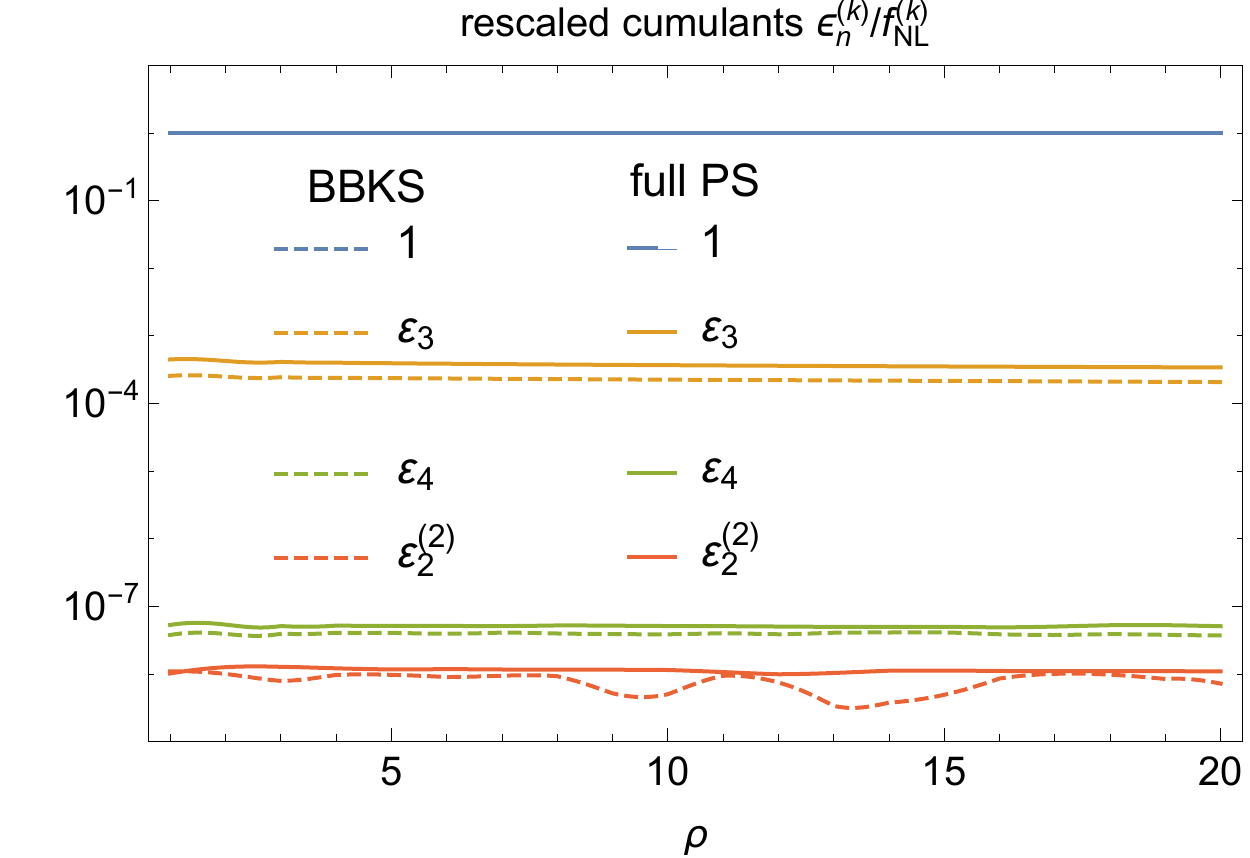}
\caption{The scale dependence of the rescaled cumulants $\varepsilon_n=\kappa_n/\sigma^n$ divided by the appropriate power of the amplitude of primordial non-Gaussianity $f_{\rm NL}$ in the relevant region for different radii that are relevant for typical densities in spheres of radii $R\simeq 10-15$ Mpc$/h$.}  
\label{fig:cumulantspNG}
\end{figure}
 The corresponding kurtosis is
\begin{align}
\label{eq:kurtosis}
 \kappa_4(R)&\equiv\langle \delta_{R,\rm NG}^4\rangle_c \\
\notag &\simeq \frac{48 f_{\rm NL}^2}{(2\pi)^9} \! \int \! \dd^3 k_1\, \alpha_R(k_1) P_\Phi(k_1) \! \int \! \dd^3 k_2 \, \alpha_R(k_2) P_\Phi(k_2)\\
 & \! \int \! \dd^3 k_3\, \alpha_R(k_3)P_\Phi(k_{13}) \alpha_R\left(k_{123}\right)\,,\\
\notag & =\frac{3 f_{\rm NL}^2}{2\pi^6} \! \int \! \dd k_1 \, k_1^2\alpha_R(k_1) P_\Phi(k_1) \! \int \! \dd k_2 \, k_2^2\alpha_R(k_2) P_\Phi(k_2)\\
\notag & \! \int \! \dd k_3\, k_3^2 \alpha_R(k_3) \int_{-1}^1 \dd\mu_{13} \, P_\Phi(k_{13})  \int_{-1}^1 \dd\mu_{132} \, \alpha_R\left(k_{123}\right)\,,
\end{align}
where $k_{13}=|\vk_1+\vk_3|=\sqrt{k_1^2+k_3^2+2k_1k_3\mu_{13}}$ and $k_{123}=|\vk_1+\vk_2+\vk_3|=k_{13}^2+k_2^2+2k_{13}k_2\mu_{123}$.
Note that at next-to-leading order also the variance receives a contribution from primordial non-Gaussianity
\begin{align}
\label{eq:variancepNG}
\sigma^2_{\rm NG}(R) &\equiv \kappa_{2}(R) \equiv \langle \delta_{R,\rm NG}^2\rangle \\
\notag &\simeq \sigma^2_{\rm G} (R) + \frac{f_{\rm NL}^2}{2\pi^4} \! \int \! \dd k_1\, k_1^2\, P_\Phi(k_1) \! \int \! \dd k_2 \,k_2^2\, P_\Phi(k_2)\\
\notag &\qquad \qquad \quad \times \int_{-1}^1\dd\mu\, \alpha_R\left(\sqrt{k_1^2+k_2^2 + 2k_1k_2\mu}\right)^2\,.
\end{align}

\cite{RothPorciani12} remark that nonzero $f_{\rm NL}$ and $g_{\rm NL}$ might cancel for the scale-dependent bias case and that the assumption of a one-parameter model (setting either $f_{\rm NL}$ or $g_{\rm NL}$ to 0) can significantly bias the estimation of the pNG parameters when both $f_{\rm NL}$ and $g_{\rm NL}$ do not vanish and the typically expected ordering $g_{\rm NL}\simeq f_{\rm NL}^2$ does not hold.

Let us now  consider the effect of the cubic term in the local transformation~\eqref{eq:lpnG} that is parametrized by $g_{\rm NL}$ which is usually assumed to be of second order $g_{\rm NL}\simeq f_{\rm NL}^2$ but can be generated independently of $f_{\rm NL}$ \citep[see e.g.][]{BernardeauUzan03}. Again, such a cubic term can be accounted for by the modifications it induces in the lowest-order cumulants, see \cite{Matarrese00}. The sub-leading term of the variance would in fact be modified by the addition of a $g_{\rm NL}\Phi^3$ term but is cancelled by the extra term included in the bracket in equation~\ref{eq:lpnGwithgnl}. The sub-leading skewness would get an extra contribution which is however of third order and hence will be ignored. The kurtosis is already modified to leading order, gaining the extra piece
\begin{align}
\kappa_4^{f_{\rm NL},g_{\rm NL}} = \kappa_4^{f_{\rm NL}} + \frac{1}{2} \frac{g_{\rm NL}}{f_{\rm NL}^2} \kappa_4^{f_{\rm NL}}
\end{align}
which still has the nice feature of appearing as a renormalization of the previously calculated leading-order kurtosis.
For two-point statistics, the two joint kurtosis terms $\kappa_{22}$ and $\kappa_{13}$ are needed, given by the formulas (53)-(54) in \cite{Silk11} and reducing to (A5)-(A6) in \cite{DesjacquesSeljak10gnl} for pure $g_{\rm NL}$. The full expressions read
\begin{align}
\label{eq:jointkurtosis31}
\kappa_{31}(r) &= \langle\delta_{R,\rm NG}(\vx)^3\delta_{R,\rm NG}(\vx+\vr)\rangle_c\\
\notag&\simeq\frac{6}{(2\pi)^9}\prod_i \left(\int dk_i\, k_i^2 \alpha(k_i)P_\Phi(k_i)\int_{-1}^1 d\mu_i\int_0^{2\pi} d\phi_i \right)\\
\notag&\qquad\qquad\qquad \alpha(k_{123}) \exp[ir(k_1\mu_1+k_2\mu_2+k_3\mu_3)]\notag \\
 &\qquad \Bigg[g_{\rm NL} \left(1+\frac{P_\Phi(k_{123})}{P_\Phi(k_3)}\right)\notag\\
 &\quad + 4f_{\rm NL}^2 \frac{P_\Phi(k_{12})}{P_\Phi(k_2)}\left(1+\frac{P_\Phi(k_2)P_\Phi(k_{123})}{P_\Phi(k_1)P_\Phi(k_3)}\right)\Bigg]
\end{align}
and
\begin{align}
\label{eq:jointkurtosis22}
\kappa_{22}(r) &= \langle\delta_{R,\rm NG}(\vx)^2\delta_{R,\rm NG}(\vx+\vr)^2\rangle_c \\
\notag &\simeq\frac{2}{(2\pi)^9}\prod_i  \left(\int dk_i \, k_i^2 \alpha(k_i)P_\Phi(k_i)\int_{-1}^1 d\mu_i\int_0^{2\pi} d\phi_i\right)\notag\\
\notag&\qquad\qquad\qquad \alpha(k_{123}) \exp[ir(k_1\mu_1+k_2\mu_2)]\notag\\
&\quad \Bigg\{3g_{\rm NL} \left(1+2\frac{P_\Phi(k_{123})}{P_\Phi(k_1)}+\frac{P_\Phi(k_{123})}{P_\Phi(k_3)}\right)\notag \\
&\quad + 4f_{\rm NL}^2 \Bigg[\frac{P_\Phi(k_{13})}{P_\Phi(k_3)}\left(1+\frac{P_\Phi(k_3)P_\Phi(k_{123})}{P_\Phi(k_1)P_\Phi(k_2)}\right)\notag\\
\notag &\qquad+\frac{P_\Phi(k_{12})+P_\Phi(k_{23})}{P_\Phi(k_2)} \left(1+\frac{P_\Phi(k_2)P_\Phi(k_{123})}{P_\Phi(k_1)P_\Phi(k_3)}\right)\Bigg]\Bigg\}
\end{align}
with $\mu_i=\cos\angle(\vr,\vk_i)$ and $k_{123}=|\vk_1+\vk_2+\vk_3|=(k_1^2+k_2^2+k_3^2+2k_1k_2\mu_{12}+2k_1k_3\mu_{13}+2k_2k_3\mu_{23})^{1/2}$ where $\mu_{i,j}=\cos\angle(\vk_i,\vk_j)=
[(1 - \mu_i^2) (1 - \mu_j^2) \cos(\phi_i - \phi_j) + \mu_i\mu_j]^{1/2}$.

\subsection{Edgeworth expansion}
\label{app:Edgeworth}

\subsubsection{Univariate Edgeworth expansion for one-point PDF}
\label{app:EdgeworthPDF}

Since primordial non-Gaussianity is small, one typically systematically expands the initial PDF of the smoothed density field in an Edgeworth expansion \cite{BernardeauKofman94,Juszkiewicz95} around a Gaussian distribution. The Edgeworth series $E_n$ is an asymptotic expansion to approximate a probability distribution using its cumulants $\kappa_n$. With the Gaussian distribution as reference function it can be written as, see Eq.\,(43) in \cite{Blinnikov97},
\begin{subequations}
\begin{align}
\label{eq:Edgeworth}
 E_n(x)&=\frac{1}{\sqrt{2\pi\kappa_2}}\exp\left(-\frac{(x-\kappa_1)^2}{2\kappa_2}\right)\\
\notag & \times \Bigg[1+ \sum_{s = 1}^n \sum_{r = 1}^s \frac{B_{s,r}(\lambda_3,..., \lambda_{s-r+3})}{s!}  H_{s+ 2 r}\left(\frac{x-\kappa_1}{\sqrt {\kappa_2}}\right) \Bigg] \,,
\end{align}
where $\lambda_n$ are the normalized and rescaled cumulants
\begin{equation}
\lambda_n \equiv \frac{\varepsilon_n}{n(n-1)} \,,
\end{equation}
$B_{s,r}$ the Bell polynomials 
and $H_n$ the probabilists' Hermite polynomials. Up to second order, the relevant Bell polynomials are
\begin{align}
\label{Bell}
B_{1,1}(\lambda_3)=\lambda_3 \ , \
B_{2,1}(\lambda_3,\lambda_4)=\lambda_4 \ ,\
B_{2,2}(\lambda_3)=\lambda_3^2 \,,
\end{align}
and the higher order Hermite polynomials read
\begin{align}
\label{Hermite}
H_4(x)&=x^4-6x^2+3\,,\\
\notag H_6(x)&=x^6-15x^4+45x^2-15\,.
\end{align}
\end{subequations}
\balance
When the Edgeworth expansion is evaluated up to next-to-leading order and expressed through the reduced primordial cumulants $\tilde S_n$ the PDF reads 
\begin{align}
\label{eq:PDFEdgeworth}
\mP(\tau) &= \frac{1}{\sqrt{2\pi}\sigma(r)} \exp \left(-\frac{\tau^2}{2\sigma^2(r)}\right) \\
&\times \Bigg[1+ \frac{\varepsilon_3(r)}{3!} H_3\left(\frac{\tau}{\sigma(r)}\right)\notag \\
\notag &\qquad \quad + \frac{\varepsilon_4(r)}{4!}H_4\left(\frac{\tau}{\sigma(r)}\right)+\frac{\varepsilon_3(r)^2}{72}H_6\left(\frac{\tau}{\sigma(r)}\right)\Bigg]\,.
\end{align}
While the Edgeworth expansion is usually accurate in the region that is close to the peak, it leads to bigger deviations in the tails. Indeed, at least to next-to-leading order the Edgeworth expansion can be obtained from the saddle-point approximation by expanding the exponential with the non-quadratic part of the rate function given in equation~\ref{eq:ratefctpNGsaddle}. Note that the modification of the variance at second order should in principle be included here by further expanding $1/\sigma^2(r)=(1-\varepsilon_2(r))/{\sigma_G^2(r)}$ in the Gaussian term when the next to leading order in $f_{\rm NL}$, or the leading order $g_{\rm NL}$), is considered, but in practice turns out to be further suppressed, see Figure~\ref{fig:cumulantspNG}.

\subsubsection{Bivariate Edgeworth expansion for two-point sphere bias}
\label{app:Edgeworthbias}
The extension of the bivariate Edgeworth expansion for the two-point PDF from the leading order given in equation~\eqref{eq:bivariateEdgeworth1st} to next-to-leading order is
\begin{align*}
\mP(\nu,\zeta) &\approx \frac{\exp\left(-\frac{\nu^2}{2} - \frac{\zeta^2}{2}\right) }{2\pi} 
 \Bigg[ 1  \!+\! \frac{1}{3!} \Big(\epsilon_{30} H_3(\nu) + \epsilon_{03} H_3(\zeta) \Big)\notag \\ 
&+ \frac{1}{2} \Big(\epsilon_{12} H_1(\nu) H_2(\zeta) 
+ \epsilon_{21} H_2(\nu) H_1(\zeta) \Big) 
\\
&+\frac{1}{4!} \Big(\epsilon_{40} H_4(\nu)  +\epsilon_{04} H_4(\zeta) \Big) + \frac{1}{4} \epsilon_{22} H_2(\nu)H_2(\zeta)\notag\\
&+\frac{1}{3!} \Big(\epsilon_{31} H_3(\nu)H_1(\zeta) + \epsilon_{13} H_1(\nu)H_3(\zeta)\Big)\notag \\
&+ \frac{1}{72} \Big(\epsilon_{30}^2 H_6(\nu)+\epsilon_{03}^2  H_6(\zeta)\Big) +\frac{1}{4} \epsilon_{12}\epsilon_{21}H_3(\nu)H_3(\zeta) \\
&+\frac{1}{12}  \Big(\epsilon_{30}\epsilon_{21} H_5(\nu)H_1(\zeta)+\epsilon_{12}\epsilon_{03}H_1(\nu)H_5(\zeta)\\
&\qquad +\epsilon_{30}\epsilon_{12} H_4(\nu)H_2(\zeta)+\epsilon_{21}\epsilon_{03}H_2(\nu)H_4(\zeta)\Big)\\
&+\frac{1}{8} \Big(\epsilon_{21}^2H_4(\nu)H_2(\zeta) + \epsilon_{12}H_2(\nu)H_4(\zeta)\Big)
\Bigg].
\end{align*} 
 using the notation $\epsilon_{ij}=\langle\nu^i\zeta^j\rangle_c$ for the joint cumulants.
 
Note that, the initial sphere bias for a pure $g_{\rm NL}$ model where $\kappa_3=\kappa_{12}=0$, in complement to equation~\eqref{eq:KaiserbiasnonGini} for pure $f_{\rm NL}$, is given as
\begin{align}
\label{eq:KaiserbiasnonGinignl}
b^{\rm NG, g_{\rm NL}}_{\rm ini}(\tau,r) &= \frac{\tau}{\sigma^{2}} \left\{1+\frac{1}{6} \left[\left(\frac{\tau}{\sigma}\right)^2-3\right] \left(\frac{\kappa_{13}(r)}{\sigma^2\xi(r)} - \frac{\kappa_4}{\sigma^4}\right) \right\} \,.
\end{align}
Interestingly, $g_{\rm NL}$ predicts an antisymmetric signature of the sphere bias for high and low densities, meaning that the sign of the sphere bias changes when comparing over- and underdensities. This is in contrast to the result for $f_{\rm NL}$ from equation~\eqref{eq:KaiserbiasnonGini} which has the same sign for both high and low densities. Hence, sphere bias for both high and low densities is promising to disentangle $f_{\rm NL}$ and $g_{\rm NL}$, which appear degenerate when looking halo bias alone.
\end{document}